\documentstyle[titlepage,12pt,epsfig,psfig]{article}
\newcommand{\Li}[1]{\mbox{Li}_2\left(#1\right)}
\newcommand{\bea}{\begin{eqnarray}}
\newcommand{\eea}{\end{eqnarray}}
\newcommand{\be}{\begin{equation}}
\newcommand{\ee}{\end{equation}}

\newcommand{\spincolorsum}{\sum_{\rm colors \atop q\bar{q}g-spins}}
\newcommand{\spinsum}{\sum_{\rm q\bar{q}g-spins \atop \ }}
\newcommand{\msbar}{{\overline{\mbox{MS}}}}
\newcommand{\xb}{\bar x}
\newcommand{\pslash}{p\!\!\!/}
\newcommand{\tr}{\mbox{Tr}}
\renewcommand{\Re}{\mbox{Re}}
\renewcommand{\Im}{\mbox{Im}}

\renewcommand{\thefootnote}{\fnsymbol{footnote}}

\makeatletter

\begin{document}

%
%
\begin{titlepage}
\noindent
hep-ph/9708350 \hfill
PITHA 97/29
\vspace{0.8cm}
\begin{center}
{\bf\LARGE 
Next-to-Leading Order QCD Corrections  \\}
{\bf\LARGE and Massive Quarks in 
$e^+e^-\to 3$ Jets\footnote{supported  
by BMBF, contract 057AC9EP.}\\}
\vspace{2cm}
\centerline{Arnd Brandenburg\footnote{Research supported by
Deutsche Forschungsgemeinschaft}$^,$\footnote{
{\tt e-mail: arndb@physik.rwth-aachen.de}} and Peter Uwer\footnote{
{\tt e-mail: uwer@physik.rwth-aachen.de}}}
\vspace{1cm}
\centerline{Institut f\"ur Theoretische Physik,
RWTH Aachen, D-52056 Aachen, Germany}
\date{\today}
\vspace{3cm}
{\bf Abstract:}\\[2mm]
\parbox[t]{\textwidth}
{We present in detail a calculation of the  
next-to-leading order QCD corrections
to the process $e^+e^-\to 3$ jets with massive quarks.
To isolate the soft and collinear divergencies of the four parton matrix
elements, we modify the phase space slicing method to account
for masses.  
Our computation allows for the prediction of oriented
three jet events involving heavy quarks, 
both on and off the
Z resonance, and of any event shape variable which is dominated
by three jet configurations. 
We show next-to-leading order results for
the three jet fraction,  
the differential two jet rate, and for the
thrust distribution at various c.m. energies.}
\end{center}
\end{titlepage}
\newpage

\setcounter{footnote}{0}
\renewcommand{\thefootnote}{\arabic{footnote}}
%
%

\begin{section}{Introduction}
High energy $e^+e^-$ annihilation experiments have proven 
to be a particularly clean testing ground for 
quantum chromodynamics (QCD), the theory of strong
interactions.  One of the key predictions  of QCD has been
the occurence of jets of hadrons in high energy
reactions  \cite{ElGaRo76,StWe77}. 
Quantitatively,
the  observed hadronic events can be classified according 
to the number of jets. The most commonly used definitions
of jets in $e^+e^-$ annihilation are based on iterative 
jet clustering algorithms. One begins with a set of
final-state particles, and considers for each iteration
the pair $\{i,j\}$ with the smallest value of a dimensionless
measure $y_{ij}$. 
If this $y_{ij}$ is smaller than a preset
cutoff $y_{cut}$, the pair is clustered into a single
pseudoparticle  according to a recombination rule. 
The procedure
is repeated until $y_{ij}>y_{cut}$ for all pairs of 
(pseudo-)particles, at which point the remaining objects 
are called jets. In the perturbative calculation,
the quantity $y_{ij}$ is constructed from the parton momenta. 
\par 
There exist a number of jet observables that are well defined
in QCD, and which can be calculated
perturbatively as an expansion in the strong coupling
$\alpha_s$. It is also possible to define a variety of 
quantities (so-called event shape variables) that
depend on the final state structure and do
not need the above concept of a jet. One of the
well-known event shape variables is thrust \cite{Fa77}.
The only requirement 
for such quantities is that they are {\it infrared safe}, 
i.e. they neither distinguish between a final state 
with one particle being soft and a final state where
this particle is absent, nor between a final state
with two collinear partons and a final state
where these two collinear partons are clustered
into a single hard pseudoparticle.   
Infrared safe quantities
that are dominated by three jet configurations, 
and the three jet differential cross section 
(defined by the jet algorithm described above) are of interest,
in particular  because
these quantities are in leading order (LO) proportional
to $\alpha_s$.  
The next-to-leading order (NLO)
QCD corrections to the production of three partons 
were computed more than fifteen years
ago \cite{ElRoTe81,Fa81,Ve81,Ku81} for massless quarks. 
Subsequently, these results have been implemented
in numerical programs \cite{KrLa89}--\cite{CaSe96}
and widely used for tests of QCD with
jet physics. Also, a complete calculation for oriented
three jet events with massless quarks was performed 
in \cite{KoSc85,KoSc89}. 
\par
To date huge samples of jet events produced at the 
$Z$ resonance have been
collected both at LEP and SLC. From these data
large numbers of jet events 
involving $b$ quarks can be
isolated with high purity using vertex detectors. 
While the mass of the $b$ quark can be neglected to 
good approximation
in the computation of the total $b\bar{b}$ 
production rate at the $Z$ peak, the
n-jet cross sections depend, 
apart from the c.m. energy $\sqrt{s}$, on an extra scale,
namely the jet resolution parameter $y_{cut}$. 
For small $y_{cut}$ 
effects due to quark masses can be enhanced.
It is therefore desirable for precision tests of 
QCD to use NLO  matrix elements
that include the full quark mass dependence.
One can then also try to extract 
the mass of the $b$ quark from three jet rates involving $b$ quarks
at the $Z$ peak, as
suggested in \cite{BiRoSa95}, elaborated in 
\cite{Ro96}, \cite{BiRoSa97}, and experimentally pursued
by the DELPHI collaboration \cite{Fu97}.
Further applications include precision tests of the 
asymptotic freedom property
of QCD by means of three jet rates and event shape variables
measured at various
center-of-mass energies, also far below the $Z$ resonance 
\cite{Be97}. For theoretical predictions 
concerning the production of top quark pairs at a future
$e^+e^-$ collider, the inclusion of the full mass dependence
is of course mandatory.
\par
The three, four, and five jet rates involving massive quarks
have been computed to leading order in $\alpha_s$ already some
time ago \cite{Jo78,Al80}. In reference \cite{MaMa96},
the NLO corrections to the production of a heavy quark
pair plus a hard photon are given.
Recently, results for the decay rate of the $Z$ boson and
the $e^+e^-$ annihilation cross section at the $Z$ peak 
into three jets including mass effects at NLO have been reported
\cite{Ro96}, \cite{BeBrUw97}, \cite{BiRoSa97}. 
Further, momentum correlations in $Z \to b\bar{b}X$ at NLO
have been studied \cite{NaOl97}. 
\par 
We have computed the complete differential
distributions for $e^+e^-$ annihilation into three and four partons
via a virtual photon or $Z$ boson at order $\alpha_s^2$,
including the full quark mass dependence. 
This allows for order $\alpha_s^2$
predictions  
of oriented three jet events, and 
of any quantity that gets contributions only 
from three and four jet configurations.
In this article, we would like to discuss this calculation in 
detail\footnote{A short account of our work is given 
in \cite{BeBrUw97}.}. We start in  section \ref{outline} with an
overview of the calculation. Section \ref{kinematics} contains
a general kinematical analysis and  the  LO results for
$e^+e^-\to 3$ jets including mass effects.  
The virtual corrections
are discussed in section \ref{virtual}. They contain singularities
that cancel against real soft and collinear contributions from four
parton final states. To isolate the latter contributions, a  
modification of the phase space slicing 
method incorporating massive partons is developed in section \ref{slicing}. 
In section \ref{numerics}
we show  numerical results. In particular, we compare for some
observables the massive with the massless result. We summarize our 
results in section \ref{concl}. An Appendix contains the analytical
result for the virtual corrections to the three parton
production rate.  
\end{section}
\begin{section}{Outline of the calculation}
\setcounter{equation}{0}
\label{outline}
The calculation of an arbitrary quantity  
dominated by three jet configurations  
and involving a massive quark-antiquark pair 
to order $\alpha_s^2$ proceeds as follows:
\par 
We first have to compute the fully differential cross
section for the partonic reaction
\bea
\label{reac1}
e^+e^- \to \gamma^\ast,Z^\ast \to Q\bar{Q}g
\eea
at leading and next-to-leading order in $\alpha_s$. Here
$Q$ denotes a massive quark and $g$ a gluon. 
\par
At leading order, the parton momenta can be identified with the jet momenta.
If the identity
of the particles making up the jets is known,
we have for the differential three jet cross section
\bea \label{jet1}
d\sigma_1(e^+e^-\to 3\ {\rm jets}) =
\Theta(y_{Q\bar{Q}}-y_{cut})\Theta(y_{Qg}-y_{cut})
\Theta(y_{\bar{Q}g}-y_{cut})
d\sigma_1(e^+e^-\to Q\bar{Q}g),
\eea
where the lower index 1  means the result in order 
$\alpha_s^1$, $\Theta$ is the Heaviside step function,
$y_{cut}$ is the jet resolution parameter
and $y_{ij}\in\{y_{Q\bar{Q}},y_{Qg},y_{\bar{Q}g}\}$ 
is determined by the experimental jet
definition. For example
\bea  \label{JaDu}
y_{ij}\ &=&\  \frac{2E_iE_j}{s}(1-\cos\theta_{ij})\ \ \ \ \ 
{\mbox{ for the JADE algorithm \cite{Ba86}}}, \nonumber \\ 
y_{ij} \ &=&\  \frac{2\ {\mbox{min}}\{E_i^2,E_j^2\}}{s}
(1-\cos\theta_{ij})\ \ \ \ \ 
{\mbox{ for the Durham algorithm \cite{Br90}}},
\eea
where $\theta_{ij}$ is the angle between jet $i$ and $j$
in the c.m. system.
At higher order and in the experimental analysis, these definitions
are supplemented by a recombination rule to form a pseudoparticle $k$
from a clustered pair $\{i,j\}$. For the above algorithms, one simply
adds the four-momenta, $p_k=p_i+p_j$. 
(For a detailed discussion of these and other jet algorithms, see also
\cite{BeKu92}.) 
If the particle content of one or more jets 
is {\it not}  
known, the LO calculation
of a three-jet quantity from 
$d\sigma_1(e^+e^-\to Q\bar{Q}g)$ may involve an averaging
over the different ways of assigning jet momenta to parton
momenta.  
\par
The one-loop integrals,
which appear in the NLO amplitude for (\ref{reac1}), contain 
ultraviolet (UV) and  
infrared (soft and collinear) (IR) singularities. 
In the
framework of dimensional regularization in $d=4-2\epsilon$
space-time dimensions, these singularities appear as poles
in $\epsilon$. The UV singularities are removed by
renormalization. Details will be given in section \ref{virtual}.
\par
After renormalization, the virtual corrections to 
the differential cross section for (\ref{reac1})
still contain IR singularities. These have to be cancelled 
by the singularities that are obtained upon 
phase space 
integration of the squared tree amplitudes for the production 
of four partons.
The relevant processes are
\bea
\label{reac2}
& & e^+e^- \to \gamma^\ast,Z^\ast \to Q\bar{Q}gg, \nonumber \\
& & e^+e^- \to \gamma^\ast,Z^\ast \to Q\bar{Q}q\bar{q}, \nonumber \\
& & e^+e^- \to \gamma^\ast,Z^\ast \to Q\bar{Q} Q\bar{Q},
\eea
where $q$ denotes a light (massless) quark. 
The first two reactions of (\ref{reac2}) give
singular contributions in the three jet region
when a parton becomes soft and/or two partons become
collinear. 
The cancellation of the singularities  has 
of course to be performed analytically. To this end, the soft and 
collinear poles of the relevant four parton matrix elements have 
to be explicitly 
isolated. We do this by modifying the so-called phase space
slicing method \cite{GiGl92} to account for masses. (For alternative
methods, see, e.g., \cite{FrKuSi96,NaTr97,CaSe96} and references therein.)
\par
The idea of the phase space slicing method is to introduce an
unphysical  parameter $s_{min}$,
which is much smaller than all  
relevant physical scales of the problem, e.g. 
$s_{min}\ll sy_{cut}$ for jet cross sections defined
by the jet resolution $y_{cut}$. The parameter $s_{min}$ 
splits the four parton
phase space into a region where all partons are ``resolved'' and
a region where at least one parton is soft and/or two partons 
are collinear. In the massless case, a convenient definition
of the resolved region is given by the requirement
$s_{ij}>s_{min}$ for all invariants $s_{ij}=(p_i+p_j)^2$.
We will modify this definition  to account for masses, see
section \ref{slicing}, but still will use the terminology
``resolved'' and ``unresolved'' partons.  
In the regions with unresolved partons, 
soft and collinear approximations 
of the matrix elements, which hold exactly in the limit
$s_{min}\to 0$, are used. The necessary integrations over
the soft and collinear regions of phase space  
can then be carried out 
analytically in $d$ dimensions.
One can thus isolate all the poles in $\epsilon$
and perform the cancellation of the IR singularities between
real and virtual contributions, after which one takes the limit 
$\epsilon\to 0$ (cf. eq. (\ref{3resolved})).
The result is a finite differential cross section 
$d\sigma_2^R(e^+e^-\to Q\bar{Q}g)$  at order $\alpha_s^2$ for 
three resolved partons, which depends on $s_{min}$. 
Its contribution to a given 
three-jet quantity is specified
by the experimental jet definition, which may involve
an energy ordering, a tagging prescription, etc.
\par
The contribution to a three jet quantity 
or to an event shape variable of 
the resolved part of the four parton cross sections
 $d\sigma_2^R(e^+e^-\to 4\ {\rm partons})$ is 
finite by itself
and may be evaluated in $d=4$ dimensions, which greatly
simplifies the algebra.
Its contribution to
three jet configurations also  depends on $s_{min}$ and is
obtained by recombining the four resolved partons into
three jets according to some recombination scheme 
and by projecting
the four parton phase space onto the experimentally specified
three jet phase space. 
\par 
Since the parameter $s_{min}$ is
introduced in the theoretical calculation
for technical reasons only and unrelated 
to any physical quantity, the sum of all contributions 
to any three jet observable must not depend on $s_{min}$.
In the soft and collinear approximations 
one neglects terms which vanish
as $s_{min}\to 0$. This limit can  be carried out numerically.
Since the individual
contributions depend logarithmically
on $s_{min}$, it is a nontrivial test of the
calculation to demonstrate that the sum of resolved and unresolved
contributions
becomes independent of
$s_{min}$ for small values
of this parameter. Moreover, in order to avoid large numerical
cancellations, one should determine the largest value of $s_{min}$
which has this property.
\par
Schematically, we have for the differential three jet
cross section involving massive quarks at order $\alpha_s^2$: 
\bea \label{resolved}
d\sigma_2(e^+e^-\to 3\ {\rm jets}) \ &=& \ 
{\bf{\Theta}} 
\bigg\{
d\sigma_2^R(e^+e^-\to Q\bar{Q}g) 
\nonumber \\ \ &+& \ 
\int
\bigg[d\sigma_2^R(e^+e^-\to Q\bar{Q}gg)+
d\sigma_2^R(e^+e^-\to Q\bar{Q}q\bar{q})
\nonumber \\ \ &+&\ 
d\sigma_2^R(e^+e^-\to Q\bar{Q}Q\bar{Q})
\bigg]\bigg\},
\eea 

where ${\bf{\Theta}}$ contains the 
experimental jet definition,    
and the 
 integration symbol 
contains the procedure of recombining resolved partons
into jets  and  of projecting  the four parton phase space 
onto the three jet phase space.
Eq. (\ref{resolved}) is schematic in the sense that
the exact relation between a given three jet quantity and
the parton cross sections depends on
the experimental knowledge about the jets, as already
mentioned above. 
 
\par 
It should be emphasized
that the notion of resolved/unresolved partons 
is unrelated to the physical jet resolution 
criterium or to any other 
relevant physical scale. In particular, we will use the freedom
in defining the soft and collinear regions of phase space
to simplify the necessary analytic integrations in $d$ dimensions.
Explicit expressions for the resolved cross sections entering
(\ref{resolved}) will be derived in section \ref{slicing}.
\par
The fully differential cross section for 
the production of {\it two} jets with massive quarks
is known to order $\alpha_s$ already for some time
\cite{JeLaZe82}. For order $\alpha_s^2$
calculations of quantities that also get contributions 
from two jet configurations, the  two 
loop amplitude for $e^+e^-\to Q\bar{Q}$ is needed.
Results for the total cross section for $e^+e^-\to$ hadrons  
to order $\alpha_s^2$ with mass corrections are reviewed
in \cite{ChKuKw96}. (For  
recent results see also \cite{ChKuSt97}.)
Since 
\bea
\sigma(e^+e^-\to {\mbox{hadrons}})&=&
\sigma(e^+e^-\to {\mbox{2 jets}})
+\sigma(e^+e^-\to {\mbox{3 jets}}) \nonumber \\
&+& \sigma(e^+e^-\to {\mbox{4 jets}})+O(\alpha_s^3),
\eea
one can now also calculate the two jet cross section
to this order, but not yet arbitrary
differential distributions.
\end{section}
%
%
\begin{section}{Kinematics and leading order results}
  \label{kinematics}
\setcounter{equation}{0}
  In this section we review some of the basic kinematics relevant for the 
  process 
  \begin{equation}
    \label{reaction}
    e^+(p_+)e^-(p_-)\rightarrow Z^\ast,\gamma^\ast\rightarrow Q(k_1) \bar Q(k_2) 
  g(k_3)
  \end{equation}
and give the leading order results for the fully differential
cross section for (\ref{reaction}).
The amplitude 
  ${\cal T}$ for the reaction (\ref{reaction}) (see also  
  Fig. \ref{fig:amplitude})
  can be written, to  order $\alpha_s^{3/2}$ and to leading order
in the electroweak couplings  in the general form:
  \begin{eqnarray}
    \label{amplitude}
    {\cal T} &=& \frac{4\pi\alpha}{s}\Big\{\,\chi(s)\, 
     \bar v(p_+)
      (g_v^e  \gamma_{\mu} - g_a^e 
    \gamma_{\mu}\gamma_5) u(p_-)\,
    (g_v^Q V^\mu - g_a^Q A^\mu - \sum_i g_a^i A_i^{\mu} )\nonumber\\
    &+& \bar v(p_+) \gamma_{\mu} u(p_-) (- Q_Q V^{\mu})\Big\}.
  \end{eqnarray}
In (\ref{amplitude}), $s=(p_++p_-)^2$, $Q_Q$ denotes 
the electric charge of the quark
in units of $e=\sqrt{4\pi\alpha}$, and 
  $g_v^f$, $g_a^f$ are the vector- and the axial-vector couplings of a
  fermion of type $f$, i.e.
  \begin{displaymath}
    g_v^f = T_3^f - 2\, Q_f\,\sin^2\vartheta_W, \quad
    \mbox{and} \quad g_a^e= T_3^f.
  \end{displaymath}
  In particular, $g_v^e = -\frac{1}{2} + 2 \sin^2\vartheta_W$, 
  $g_a^e =-\frac{1}{2}$ for an electron or  
  $g_v^b = -\frac{1}{2} + \frac{2}{3} \sin^2\vartheta_W$,
  $g_a^b =-\frac{1}{2}$ for a bottom quark, with $\vartheta_W$ denoting the 
  weak mixing angle.
  The function $\chi(s)$ is given by
  \begin{equation}
    \label{chi}
    \chi(s) = \frac{1}{4\sin^2\vartheta_W\cos^2\vartheta_W}\,
    \frac{s}{s-m_Z^2 + i m_Z \Gamma_Z},
  \end{equation}
  where $m_Z$ and $\Gamma_Z$ stand for the mass and the width of the Z-boson.
  We have defined in (\ref{amplitude}) 
  the amplitudes $V^{\mu}, \ A^{\mu}$, and $A^\mu_i$, which contain the
  information on the decay of the vector boson into the three final state
  partons.
  The contribution $A^\mu_i$  is generated at order $\alpha_s^{3/2}$
  by a quark triangle  loop \cite{HaKuYa91}.
  The sum runs over the different quark flavors. Only quark doublets with
  a mass-splitting give a non-vanishing contribution.
   At higher order
  in $\alpha_s$  there are
  additional terms in eq. (\ref{amplitude}) of the form $V_i^\mu$, which are 
absent at order $\alpha_s^{3/2}$ due to a non-abelian generalization 
of Furry's theorem. Such additional terms are 
also present in  the scattering amplitude 
for the four parton final state $q\bar{q}Q\bar{Q}$ with light quarks at the initial
vertex at lowest order in $\alpha_s$.  

   The
  squared amplitude $|{\cal T}|^2$ can be written to order 
  $\alpha_s^2$ 
  in the compact form:
  \begin{eqnarray}
    \label{master}
|{\cal T}|^2 &=& \frac{16\pi^2\alpha^2}{s^2} 
\Big[L^{PC\mu\nu}H^{PC}_{\mu\nu}+L^{PV\mu\nu}H^{PV}_{\mu\nu}\Big].
\end{eqnarray}
  We restrict ourselves here to the relevant case of 
  longitudinal polarization of electrons and/or positrons
and neglect the lepton masses. 
  The lepton tensors $L^{PC\mu\nu}$ and
  $L^{PV\mu\nu}$ take the usual form:
  \begin{eqnarray}
    \label{leptontensors}
    L^{PC\mu\nu}&=& \frac{1}{4}\tr[\pslash_+\gamma^\mu\pslash_-\gamma^\nu]
    =  p_{+}^{\mu}p_{-}^{\nu}+p_{+}^{\nu}p_{-}^{\mu}-g^{\mu\nu}p_+p_-,
    \nonumber\\
    L^{PV\mu\nu}&=& \frac{1}{4}\tr[\gamma_5\pslash_+\gamma_\mu\pslash_-
    \gamma_\nu]= -i\varepsilon^{\mu\nu}_{\ \ \rho\sigma}p_+^{\rho}p_-^{\sigma},
  \end{eqnarray}
where $\varepsilon_{0123}=+1$.
The tensors $H^{PC(PV)}_{\mu\nu}$ (which we 
  call ``parton tensors'') may be written as follows:
\bea
\label{master2}
H^{PC(PV)}_{\mu\nu} \ &=&\       
 g^{VV}_{PC(PV)} H^{VV}_{\mu\nu} + g^{AA}_{PC(PV)} H_{\mu\nu}^{AA}
      + g_{PC(PV)}^{VA_+} H_{\mu\nu}^{VA_+} 
      + \sum_i g_{PC(PV)}^{AA^i} H_{\mu\nu}^{AA^i} \nonumber \\
      &+& \sum_i g_{PC(PV)}^{VA_+^i} H_{\mu\nu}^{VA_+^i}
      + g_{PC(PV)}^{VA_-}H_{\mu\nu}^{VA_-}      
      + \sum_i g_{PC(PV)}^{VA_-^i}H_{\mu\nu}^{VA_-^i}.
\eea
The coupling constants are given explicitly by 
\begin{eqnarray}
g_{PC (PV)}^{VV} &=&Q_Q^2\, f_{PC(PV)}^{\gamma\gamma} 
    + 2 \,g_v^Q\,Q_Q\, \Re \chi(s)\,f_{PC(PV)}^{\gamma Z} 
    + g_v^{Q\,2}\, |\chi(s)|^2\,f_{PC(PV)}^{ZZ},\nonumber\\
g_{PC(PV)}^{AA}  &=& g_a^{Q\,2} |\chi(s)|^2 f_{PC(PV)}^{ZZ},
\nonumber\\
g_{PC(PV)}^{VA_+}&=& -g_a^Q\,Q_Q\,\Re\,\chi(s)\,f_{PC(PV)}^
{\gamma Z} -g_v^Q\,g_a^Q\, |\chi(s)|^2 f_{PC(PV)}^{ZZ},\nonumber\\
g_{PC(PV)}^{AA_i} &=& g_a^{Q}\,g_a^{i}|\chi(s)|^2 f_{PC(PV)}^{ZZ},
\nonumber\\
g_{PC(PV)}^{VA_+^i}&=& -g_a^i\,Q_Q\,\Re\,\chi(s)\,
f_{PC(PV)}^{\gamma Z} 
-g_v^Q\,g_a^i\, |\chi(s)|^2 f_{PC(PV)}^{ZZ},\nonumber\\
g_{PC(PV)}^{VA_-}&=&i\,g_a^Q\,Q_Q\,\Im\chi(s) f_{PC(PV)}^{\gamma Z},
\nonumber\\
g_{PC(PV)}^{VA_-^i}&=&i\,g_a^i\,Q_Q\,\Im\chi(s) 
f_{PC(PV)}^{\gamma Z},
\end{eqnarray}
where
\begin{eqnarray}
f_{PC}^{\gamma\gamma}&=&1-\lambda_-\lambda_+,\nonumber\\
f_{PC}^{\gamma Z}&=&-(1-\lambda_-\lambda_+)g_v^e + 
(\lambda_--\lambda_+)g_a^e,\nonumber\\
f_{PC}^{ZZ}&=&(1-\lambda_-\lambda_+)(g_v^{e2}+g_a^{e2})-
2(\lambda_--\lambda_+) g_v^{e} g_a^{e},\nonumber\\
f_{PV}^{\gamma\gamma}&=& \lambda_--\lambda_+,\nonumber\\
f_{PV}^{\gamma Z}&=& (1-\lambda_-\lambda_+)g_a^e -(\lambda_--\lambda_+)
g_v^e,\nonumber\\
f_{PV}^{ZZ}&=&(\lambda_--\lambda_+) (g_v^{e\,2}+g_a^{e\,2}) -
2\,(1-\lambda_-\lambda_+)g_v^e\,g_a^e,
\end{eqnarray}
with $\lambda_-$ ($\lambda_+$) denoting the longitudinal 
polarization  of the electron (positron) beam.
\par
  The parton tensors $H^{VV}_{\mu\nu}, H^{AA}_{\mu\nu},\ldots$ 
  are related to the amplitudes
  $V_\mu, A_\mu$ and $A^i_\mu$ in the following way
(we do not write down explicitly the sum over the polarizations
of the final states):
  \begin{eqnarray}
    \label{tensors}
    H_{\mu\nu}^{VV} &=& V_\mu V_\nu^\ast,\nonumber\\ 
    H_{\mu\nu}^{AA} &=& A_\mu A_\nu^\ast, \nonumber\\
    H_{\mu\nu}^{VA_\pm} &=& V_\mu A_\nu^\ast \pm A_\mu V_\nu^\ast,\nonumber\\
    H_{\mu\nu}^{VA^i_\pm} &=& V_\mu A_{i\nu}^\ast \pm A_{i\mu} 
    V_\nu^\ast,\nonumber\\
    H_{\mu\nu}^{AA^i} &=& A_\mu A_{\nu}^{i\ast} + A^i_{\mu} A_\nu^\ast.
  \end{eqnarray}
  In massless QCD  chiral symmetry implies that 
  $H^{VV}_{\mu\nu} = H^{AA}_{\mu\nu}$ and $H_{\mu\nu}^{VA_-}=0$. 
The tensors $H_{\mu\nu}^{VA_-}$  and  $H_{\mu\nu}^{VA_-^i}$ multiply factors
that are formally of higher order in the electroweak 
coupling. Their contributions to the differential cross section
for (\ref{reaction}) are negligibly small.
\par
The parton tensors $H_{\mu\nu}$ entering (\ref{master}) contain
the complete information about the orientation of the partons in 
the final state with respect to the beam axis and about their
energy distribution. 
Although the parton tensors are good for calculational 
purposes\footnote{For
the computation of $H_{\mu\nu}$ in the massless case 
to order $\alpha_s^2$, see \cite{KoSc85}.}, 
we find it more convenient for the phenomenological application 
to switch to a slightly 
different notation, which we adopt from ref. \cite{HaKuYa91}.  
We write the $e^+e^-\to 
\gamma^\ast,Z^\ast\to Q\bar{Q}g$ 
 fully differential cross section    
(in $d=4$ dimensions), with no transverse beam polarization,
as follows:
\bea
\label{parton1}
\frac{d^4\sigma}{ dx\, d\xb\, d\cos\theta\, d\phi}\ &=&\
  \frac{3}{4\pi}\frac{\alpha_s}{\pi}\sigma_{\rm pt} \Big[
F_1 (1+\cos^2\theta) + F_2 (1-3\cos^2\theta) + F_3 \cos\theta 
\nonumber \\
  &+& F_4 \sin2\theta\cos\phi + F_5 \sin^2\theta \cos2\phi
  + F_6\sin\theta\cos\phi \nonumber \\
  &+& F_7 \sin2\theta\sin\phi + F_8 \sin^2\theta \sin2\phi
  + F_9\sin\theta\sin\phi \Big]\ , 
\eea
where
\bea 
\label{ptcrosssection}
 \sigma_{\rm pt}\ =\ \sigma(e^+e^- \to \gamma^* \to \mu^+\mu^-)
  \ =\ \frac{4\pi\alpha^2}{3s}\ .
\eea

In (\ref{parton1}), $x=2kk_1/s$ and 
$\xb=2kk_2/s$ are the scaled energies of the
quark and antiquark in the $Z$ rest system, $\theta$ is the angle between the 
electron direction and the quark direction, and  $\phi$ 
is the (signed) angle between the $e^+e^-Q$ plane
and the $Q\bar{Q} g$ plane. 
The functions $F_i$ are easily obtained by 
contracting the parton tensors $H_{\mu\nu}$
with suitable tensors built from the final state
momenta and $g_{\mu\nu}$.
 \par
The functions $F_7,\ F_8$ and $F_9$ are generated by  absorptive
parts in the scattering amplitude, i.e. they are zero in
tree-approximation. In addition they
vanish for massless quarks at one loop order
\cite{KoSc80}, \cite{KoSc85}{\footnote{A nonzero function  
$F_9$ gives rise to a correlation between 
the beam axis and the normal to the three jet event plane.
This ``event handedness'' was first discussed by \cite{FaKrScSc80}
and studied in detail at the $Z$ resonance both
experimentally \cite{SLD96}
and theoretically \cite{BrDiSh96}.}}. For the nonvanishing LO functions 
$F_i$ we have:
\bea
\label{Fi01}
F_i^{(0)} &=& g_{PC}^{VV}F_i^{(0),VV}+g_{PC}^{AA}F_i^{(0),AA}\ \ \ \ \ 
(i=1,2,4,5),\nonumber \\
F_j^{(0)} &=& g_{PV}^{VA_+}F_j^{(0),VA_+}\ \ \ \ \  
(j=3,6).
\eea
Using the abbreviations 
\bea
\label{abbr}
z=\frac{m^2}{s},\ \ \ \ \  B=\frac{1}{(1-x)(1-\xb)},\ \ \ \ \ x_g=2-x-\xb,
\eea 
where $m$ is the quark mass, we find: 
\bea
\label{Fi02}
F_1^{(0),VV} &=& B\bigg\{ {x^2+\xb^2 \over 2 }
  +  z \left[-3(x+\xb)^2 + 8(x+\xb)+2x\xb(1-x_g)-6 \right]B
  - 2z^2x_g^2B\bigg\}, \nonumber \\
F_1^{(0),AA} &=& F_1^{(0),VV} + zB\bigg\{(x+\xb)^2-10(1-x_g)+6zx_g^2B\bigg\}
,\nonumber \\
F_2^{(0),VV} &=& \frac{1}{x^2-4z} \bigg\{1-x_g-z
(-3x^2+\xb^2-2x\xb+2x-2\xb+2)B \nonumber \\
&-& 2z^2[2x^3-13x^2-5\xb^2+2x\xb(4x+3\xb-11)+20x+12\xb-8]B^2
+8z^3x_g^2B^2\bigg\},\nonumber \\
F_2^{(0),AA} &=& \frac{1-x_g}{x^2-4z}-
\frac{z(1-\xb)[x^3-5x^2-3\xb^2+x\xb(2x+\xb-8)+8x+6\xb-2]
-2z^2x_g^2 }{(1-\xb)^2(x^2-4z)}
 ,\nonumber \\
F_3^{(0),VA_+} &=&\frac{2B}{\sqrt{x^2-4z}}\bigg\{x^3+2\xb^2+x\xb(-\xb+2)
-2\xb-2z[x^3-8x^2-4\xb^2\nonumber \\ &+&
x\xb(6x+5\xb-16)+12x+8\xb-4]B + 8z^2x_g^2B\bigg\},\nonumber \\
F_4^{(0),VV} &=& \frac{B\sqrt{(1-x)(1-\xb)(1-x_g)-zx_g^2}}
{(x^2-4z)(\xb^2-4z)}\bigg\{ \xb^2[2(1-x_g)-x\xb] \nonumber \\
&+&\frac{2z}{1-\xb}[4\xb^2+x\xb(3\xb^2+x\xb-6\xb+6)-4x-8\xb+4
\nonumber \\ &-& 4z(\xb^3+x^2-2\xb^2+x\xb(\xb+3)-4x)-16z^2x_g]\bigg\},
\nonumber \\
F_4^{(0),AA} &=& F_4^{(0),VV} + 
\frac{4zB\sqrt{(1-x)(1-\xb)(1-x_g)-zx_g^2}}{x^2-4z}\bigg\{x^2-2(1-x_g)+x\xb \bigg\} \nonumber \\
F_5^{(0),VV} &=& F_2^{(0),VV}
-2zB\bigg\{1-x_g-zx_g^2B\bigg\} \nonumber \\
F_5^{(0),AA} &=& F_2^{(0),AA}-zx_g^2B,\nonumber \\
F_6^{(0),VA_+} &=& 4B\sqrt
{\frac{(1-x)(1-\xb)(1-x_g)-zx_g^2}{x^2-4z}}\bigg\{\xb
-\frac{2zx_g}{1-\xb}\bigg\}
.\eea
Our results agree with the results of \cite{OlSt94} and reduce in 
the massless case to those given in \cite{HaKuYa91}{\footnote{
In reference \cite{BrDiSh96}, the expression for
$F_2^{(0),AA}-F_5^{(0),AA}$ (last equation in  (B1) of 
\cite{BrDiSh96}, denoted there $F_2^{(0),a}-F_5^{(0),a}$) 
contains a misprint: The denominator of this 
function is given there as $(1-x)^2(1-\xb)^2$, while the correct
denominator is $(1-x)(1-\xb)$, cf. eq. (\ref{Fi02}) above.}}.
The increase 
in algebraic complexity due to the non-neglection of
the quark mass is quite substantial already at LO.
%
%
\begin{section}{Virtual corrections}
  \label{virtual}
\setcounter{equation}{0}
The calculation of the virtual corrections for reaction
(\ref{reaction}) is straightforward albeit tedious. 
Non-neglection of the quark mass leads to a considerable
complication of the algebra. 
In this section we give details on this part of the calculation.
\par
We work in renormalized
perturbation theory, i.e. start with a renormalized
Lagrangian including counterterms 
to calculate the (truncated) renormalized Green function for 
reaction (\ref{reaction}) in $d$ dimensions to order
$\alpha_s^{3/2}$. 
The relevant one-loop diagrams are shown in figure \ref{fig:loop}, 
the counterterm diagrams  are depicted in figure \ref{fig:counter}.
The scattering amplitude,
\bea
{\cal T}(e^+e^-\to Q\bar{Q}g)= 
{\cal T}^{\scriptsize{\mbox{Born}}}
(e^+e^-\to Q\bar{Q}g)+
{\cal T}^{\scriptsize{\mbox{virtual}}}
(e^+e^-\to Q\bar{Q}g)+O(\alpha_s^2),
\eea
follows
from the renormalized truncated Green function
by multiplication of UV finite  
wave function renormalization
factors in accordance with the Lehmann-Symanzik-Zimmermann
reduction formalism.
Except for the quark mass renormalization, we work in the
modified minimal subtraction ($\msbar$) scheme throughout.
For the quark mass renormalization we consider both the
on-shell scheme and the $\msbar$ scheme. The relation
between both schemes is given to order $\alpha_s$ by
the well-known relation
\bea\label{massrel}
m^{\rm pole}=m^{\scriptsize{\mbox{$\msbar$}}}(\mu)
\left[1+\frac{\alpha_s(\mu)}{\pi}
\left(\frac{4}{3}-2
\ln\frac{m^{\scriptsize{\mbox{$\msbar$}}}(\mu)}{\mu}\right)\right],
\eea
where $\mu$ is the renormalization scale. It is 
somewhat more transparent to use the pole mass from
the start and, if desired, switch to the $\msbar$ scheme
only at the very end of the calculation using (\ref{massrel}).
In an {\it ab initio} calculation with the $\msbar$ mass one
has to  distinguish between the renormalized 
mass parameter  of the propagators (which then by definition
is the $\msbar$ mass) and the mass of the
external particles (which of course is always the pole mass).
\par
Throughout the calculation presented in this work the background
field gauge \cite{Ab81,AbGrSc83} is used. This gauge simplifies the 
triple-gluon vertex and in addition a simple relation 
between the renormalization 
constants of the 
coupling $Z_g$ and the wave function of the gluon field
$Z_A$  holds in the MS and $\msbar$ schemes, namely
\bea 
Z_g=Z_A^{-1/2}.
\eea
This implies also that the 
renormalization constants of the  
quark-gluon coupling $Z_{1F}$ and the quark wave function $Z_{\psi}$
are equal,
\bea
Z_{1F} = Z_{\psi}.
\eea
\par
For the calculation of the loop diagrams the Passarino-Veltman method
\cite{PaVe79} is  used to reduce the tensor 
integrals to scalar one-loop
integrals. When doing the loop integration UV as well as IR
singularities are present. Dimensional regularisation is employed
to treat both. The
formulas of \cite{PaVe79} are generalized 
in order to account for the appearance of IR  poles.
The reduction of tensor integrals and the trace algebra in
$d$ dimensions is carried out with two different programs:
{\sc Form} \cite{FORM} and {\sc Reduce}  \cite{Reduce} are used 
independently for this part of the calculation and
yield the same results. 
\par
Because of the presence of the axial-vector current, a prescription
to handle the 
$\gamma_5$ matrices in $d$ dimensions 
has to be chosen. In the case of traces with two
$\gamma_5$'s, we work with an anticommuting $\gamma_5$ in
$d$ dimensions, which is known to be consistent \cite{ChFuHi79}. 
By using the relation $\gamma_5^2=1$ 
we eliminate the $\gamma_5$'s from the traces. 
In the case of traces with only one 
$\gamma_5$ the situation is more complicated. Here we use the 
't~Hooft-Veltman prescription \cite{HoVe72}.
It is well known that this prescription
violates certain Ward identities, i.e. in the limit of vanishing 
quark masses chiral invariance is broken. 
To restore the chiral Ward identities, 
a special $\gamma_5$ counterterm  has 
to be taken into account in higher order calculations.
Explicitly, in our case
the replacement \cite{La93}
\bea 
\label{Z5}
\gamma_{\mu}\gamma_5 \to Z_5^{ns}\frac{i}{3!}\varepsilon_{\mu\nu_1
\nu_2\nu_3}\gamma^{\nu_1}\gamma^{\nu_2}\gamma^{\nu_3},\ 
{\mbox{     with }} Z_5^{ns}= 1 -\frac{\alpha_s}{\pi}C_F
\eea
restores the chiral Ward identities to order $\alpha_s^{3/2}$.
\par
When calculating the loop integrals we find it useful to replace the
box integrals in $d=4-2\epsilon$ 
dimensions by the box integrals in
$6-2\epsilon$ dimensions plus a linear combination of the
three-point integrals. This method \cite{BeDiKo93} has 
the advantage that the IR singularities 
only appear in the three-point integrals, 
because the 
box integrals in 6 dimensions are IR finite. In this way
the IR divergent part
of the virtual corrections can be easily obtained. 
Further, the coefficient functions which multiply the
loop integrals become algebraically less involved
in this way.
\par
We compute both
the real and the imaginary part of the loop integrals by
Feynman parametrization techniques. The imaginary parts
of the integrals induce several contributions 
to the differential cross
section (\ref{parton1}) : The function $F_9$ is nonzero
only through these imaginary parts, and the parity violating
functions $F_7$ and $F_8$ get the dominant contributions
from the imaginary parts of the one-loop integrals\footnote{Additional
contributions to $F_{7,8}$, which are formally of higher order in the
electroweak coupling, are induced by the 
imaginary part of the $Z$ propagator times the real parts
of the one-loop integrals  via $\gamma - Z$ interference.
While the main virtual contributions to the functions $F_3$
and $F_6$ are proportional to the real parts of the one-loop integrals,
there are additional subdominant contributions from $\gamma - Z$ interference
of the form (imaginary part of one-loop integrals)
$\times$ (imaginary part of the $Z$ propagator).}.
Our results for the imaginary parts of the integrals
agree with the results given in \cite{BrDiSh96}. 
\par
The explicit expressions for the virtual corrections 
to $F_1-F_9$ are too lengthy to be fully reproduced
here\footnote{The result for $F_9$ is given in \cite{BrDiSh96}.}.
We will restrict our discussion of the analytic results
to the function $F_1$, which
determines in particular the three parton rate,
\bea \label{parton2}
\frac{d^2\sigma}{dxd\xb}=\frac{4\alpha_s}{\pi}
\sigma_{\rm pt}F_1.
\eea 
The virtual corrections to $F_1$ may be written as
\bea
\label{F_1v}
F_1^{\scriptsize{\mbox{virtual}}}=g_{PC}^{VV}
F_1^{\scriptsize{\mbox{virtual}},VV}
+g_{PC}^{AA}F_1^{\scriptsize{\mbox{virtual}},AA}
+\sum_ig_{PC}^{AA_i}F_1^{\scriptsize{\mbox{virtual}},AA_i}.
\eea
We can  further decompose the first two terms in (\ref{F_1v})
as follows (recall that we already removed
the UV poles by renormalization; the remaining poles are due
to IR singularities):
\bea
\label{F_1v2}
F_1^{\scriptsize{\mbox{virtual}},VV(AA)}&=&
-\frac{\alpha_s}{4\pi}N_C\bigg\{\frac{2}{\epsilon^2}
+\frac{1}{\epsilon}
\bigg[\frac{17}{3}+2
\left(\ln\left(\frac{4\pi\mu^2}{s}\right)+
\ln\left(zB\right)-\gamma\right) 
\nonumber \\ &-&\frac{2n_f^{\rm ms}}{3N_C}
-\frac{1}{N_C^2}\frac{1}{\beta}(2\beta-
(1+\beta^2)\ln(\omega)
\bigg] 
\bigg\}F_1^{(0),VV(AA)} \nonumber \\ 
&+& F_1^{\scriptsize{\mbox{rest,singular}},VV(AA)} 
+ F_1^{\scriptsize{\mbox{counter,finite}},VV(AA)}+
F_1^{\scriptsize{\mbox{ext.,finite}},VV(AA)} \nonumber \\ &+&
\frac{\alpha_s}{4\pi}N_C
\left\{F_1^{{\rm lc},VV(AA)}+\frac{1}{N_C^2}F_1^{{\rm sc},VV(AA)}
\right\}
+{\cal O}(\epsilon),
\eea
where 
\bea
\label{beom}
\beta=\sqrt{1-\frac{4z}{1-x_g}},\ \ \ \ \ \omega=\frac{1+\beta}
{1-\beta},
\eea
$n_f^{\rm ms}$ is the number of massless flavors, 
the functions $F_1^{(0),VV(AA)}$ are  given in (\ref{Fi02}),
and $z$, $B$ and $x_g$ have been defined in eq. (\ref{abbr}). 
Further, 
\bea
\label{F_1v3}
F_1^{\scriptsize{\mbox{rest,singular}},VV} 
\ &=&\  \frac{1}{\epsilon}
\frac{\alpha_s}{4\pi}N_Cx_g^2B,
\nonumber \\
F_1^{\scriptsize{\mbox{rest,singular}},AA} \ &=&\
\frac{1}{\epsilon}\frac{\alpha_s}{4\pi}N_Cx_g^2B(1+2z).
\eea
The finite contributions from
the counter terms and the external wave function
factors are given in the on-shell mass renormalization scheme
for one massive and $n_f^{\rm ms}$ massless flavors by: 
\bea\label{counterext}
F_1^{\scriptsize{\mbox{counter,finite}},VV} &=& 
-\frac{\alpha_s}{\pi}B\bigg\{
\bigg[\ln(4\pi)-\gamma+\frac{4}{3}-
\ln\left(\frac{zs}{\mu^2}\right)
\bigg]
\nonumber \\ &\times&
\bigg[x^2-zB(x^3+7x^2+5x\xb(-x+2)-20x+7)
\nonumber \\ &-&
4z^2B^2(-4x^3+16x^2+2x^2\xb^2+x\xb(3x^2-17x+17)-23x+6)
\nonumber \\ &-& 8z^3x_gB^2(1-x)^2 
+ (x\leftrightarrow \xb)\bigg]
-x_g^2-zx_g^3B \bigg\}
,\nonumber \\
F_1^{\scriptsize{\mbox{counter,finite}},AA} &=& 
F_1^{\scriptsize{\mbox{counter,finite}},VV}
-\frac{\alpha_s}{\pi}zB\bigg\{
\bigg[\ln(4\pi)-\gamma+\frac{4}{3}-
\ln\left(\frac{zs}{\mu^2}\right)\bigg]
\nonumber \\ &\times&
\bigg[3(x^2-8x+x\xb+3)-2zB(x^3-24x^2+3x\xb(x-6)+66x-28)
\nonumber \\
&+& 24z^2x_gB^2(1-x)^2 
+ (x\leftrightarrow \xb)\bigg]
+3x_g(x+\xb)-2zx_g^3B\bigg\},\nonumber \\
F_1^{\scriptsize{\mbox{ext.,finite}},VV} &=&
-\frac{\alpha_s}{12\pi}\bigg\{\bigg[
(11N_C-2n_f^{\rm ms})
(\ln(4\pi)-\gamma)-2\ln\left(\frac{zs}{\mu^2}
\right)\bigg]F_1^{(0),VV}
\nonumber \\ &-& (11N_C-2n_f^{\rm ms})
\frac{x_g^2B}{2}\bigg\},
\nonumber \\
F_1^{\scriptsize{\mbox{ext.,finite}},AA} &=&
-\frac{\alpha_s}{12\pi}\bigg\{\bigg[
(11N_C-2n_f^{\rm ms})
(\ln(4\pi)-\gamma)-2\ln\left(\frac{zs}{\mu^2}
\right)\bigg]F_1^{(0),AA}
\nonumber \\ &-&
(11N_C-2n_f^{\rm ms})
\frac{x_g^2B(1+2z)}{2}\bigg\} .
\eea

The rather involved results for the finite leading-color
($F_1^{{\rm lc},VV(AA)}$)
and subleading-color ($F_1^{{\rm sc},VV(AA)}$) contributions  
from the interference of the loop diagrams, Figs. 2(a)-(k),
with the Born graphs
are listed in the appendix. 
Finally, 
\bea
\sum_ig_{PC}^{AA_i}F_1^{\scriptsize{\mbox{virtual}},AA_i} = 
g_a^Q|\chi(s)|^2f^{ZZ}_{PC}\frac{\alpha_s}{\pi}
\frac{x+\xb}{2}[1-zBx_g^2]
\sum_i g_a^i {\mbox {Re}}I^i(x,\xb,z_i),
\eea
where $z_i=m_i^2/s$ denotes the square of the scaled mass 
of the quark in the fermion triangle of Figs. 2(l),(m) and 
$\sum_i g_a^i{\mbox {Re}}I^i(x,\xb,z_i)$ can be  obtained 
by simple substitutions from formulas (2.17) and (2.18)
of \cite{HaKuYa91} and will therefore not be given 
explicitly here. This contribution to 
$F_1$ is finite by itself and numerically very small
\cite{HaKuYa91}.

\end{section}

%
%
\setcounter{equation}{0}
\bigskip
\section{The phase space slicing method for massive quarks}
\label{slicing}
 In this section we will describe how 
we modify the phase space slicing method 
in the presence of masses and derive explicit expressions for
the resolved cross sections $d\sigma^R_2$ at NLO entering
equation (\ref{resolved}).
\par
In the presence of massive quarks, the structure of
collinear and soft poles of the four parton matrix elements
is completely different as compared to the massless case.
In particular, the nonzero quark mass serves as a regularizer 
for collinear singularities. 
Thus, the matrix elements contain fewer singular structures,
but the presence of a quark mass leads to more complicated phase
space integrals.
\par
The contribution of the process 
$e^+e^-\to Q\bar{Q}Q\bar{Q}$ to the three jet
differential cross section is free of singularities. Thus, 
for this subprocess, it is possible to define:
\bea  \label{4bs}
d\sigma_2^R(e^+e^-\to Q\bar{Q}Q\bar{Q}) \ &=& \ 
d\sigma_2(e^+e^-\to Q\bar{Q}Q\bar{Q}).
\eea
The process $e^+e^-\to Q\bar{Q}q\bar{q}$ is singular
in the three jet region only when the massless partons become
collinear, whereas the contribution of 
$e^+e^-\to Q\bar{Q}gg$ contains both soft and
collinear divergencies. 
We will first discuss the 
contributions of soft gluons.
\par
The amplitude for 
${\cal T}(e^+e^-\to Q(k_1)\bar{Q}(k_2)g(k_3)g(k_4))$ 
can be  written in 
terms of color-ordered subamplitudes:
\bea
{\cal T}(e^+e^-\to Q\bar{Q}gg) = (T^{a_3}T^{a_4})_{c_1c_2}
S_1+(T^{a_4}T^{a_3})_{c_1c_2}S_2,
\eea
where $T^a$ denote color matrices, $a_3,a_4$ the color of the
gluons, and $c_1,c_2$ the color of the quarks.
For the squared matrix element 
(summed over colors and spins) we thus get
\bea \label{color2}
\spincolorsum|{\cal T}(e^+e^-\to Q\bar{Q}gg)|^2 = \frac{N_C^2-1}{2}
\frac{N_C}{2} \spinsum\left[ |S_1|^2 +|S_2|^2 - \frac{1}{N_C^2}|S_1+S_2|^2\right]
\eea
The term $|S_1+S_2|^2$ in (\ref{color2}) contains the
QED-like contributions. 
\par
In the limit where one gluon becomes soft, each of the 
terms in the squared matrix element (\ref{color2})
can be written as a factor
multiplying the squared Born
matrix element for $e^+e^-\to Q\bar{Q} g$:

\bea\label{S1lim}
\frac{N_C^2-1}{2}\frac{N_C}{2}\spinsum|S_1|^2 \ \ &
{\buildrel
k_3\to 0\over \longrightarrow}&\ \  
\frac{g_s^2N_C}{2}f_{l}(1,3,4)
\spincolorsum|{\cal T}^{\scriptsize{\mbox{Born}}}(e^+e^-\to Q\bar{Q}g)|^2,
\nonumber \\
\ \  &
{\buildrel
k_4\to 0\over \longrightarrow}&\ \ 
\frac{g_s^2N_C}{2}f_{l}(2,4,3)
\spincolorsum|{\cal T}^{\scriptsize{\mbox{Born}}}(e^+e^-\to Q\bar{Q}g)|^2.
\eea
Here $g_s$ denotes the strong coupling constant.
The limiting behavior of the term $|S_2|^2$ follows from (\ref{S1lim})
by the exchange $(1\leftrightarrow 2)$. For the term subleading in the number
of colors we have 

\bea\label{S12lim}
-\frac{N_C^2-1}{4N_C}\spinsum|S_1+S_2|^2 \ \  &
{\buildrel
k_3\to 0\over \longrightarrow}&\ \ 
-\frac{g_s^2}{2N_C}f_{sl}(1,3,2)
\spincolorsum|{\cal T}^{\scriptsize{\mbox{Born}}}(e^+e^-\to Q\bar{Q}g)|^2,
\nonumber \\
\ \  &
{\buildrel
k_4\to 0\over \longrightarrow}&\ \ 
-\frac{g_s^2}{2N_C}f_{sl}(1,4,2)
\spincolorsum|{\cal T}^{\scriptsize{\mbox{Born}}}(e^+e^-\to Q\bar{Q}g)|^2.
\eea
We have defined in (\ref{S1lim}), (\ref{S12lim}) 
 the {\it eikonal factors}
\bea \label{eikonal}
f_{l}(i,s,k)\ &=&\  \frac{4t_{ik}}{t_{is}t_{sk}}-\frac
{4m^2}{t_{is}^2},\nonumber \\
 f_{sl}(1,s,2) \ &=&\  \frac{4t_{12}}{t_{1s}t_{2s}}
-\frac{4m^2}{t_{1s}^2}-\frac{4m^2}{t_{2s}^2},
\eea
where  $t_{ij}=2k_ik_j$ and $m$ denotes the quark mass.
\par
For notational convenience, we further define
\bea 
\Theta_{isk}\ &\equiv &\  
\Theta(t_{is} + t_{sk}- 2 s_{min}),\nonumber \\
\overline{\Theta}_{isk}\ &\equiv &\  
\Theta(2 s_{min}- t_{is} - t_{sk}).
\eea
Bearing in mind the soft limits (\ref{S1lim}), let us consider
the identity
\bea \label{S1decomp}
|S_1|^2\ &=&\  
\left(\Theta_{134}+\overline{\Theta}_{134}\right)
\left(\Theta_{243}+\overline{\Theta}_{243}\right)
|S_1|^2 \nonumber \\
\ &=&\ \left(\Theta_{134}\Theta_{243} + \overline{\Theta}_{134}
+ \overline{\Theta}_{243} -  \overline{\Theta}_{134} 
\overline{\Theta}_{243}
\right)|S_1|^2.
\eea
The first term in the second line of eq. (\ref{S1decomp}) 
corresponds to the case where the energies of both gluons
are bounded from below, i.e., neither of the gluons can become 
soft. (They can still become collinear, which will be discussed
later.) The second
and the third term give rise to soft gluon singularities in
the three jet region. Finally, the last term describes 
the situation where both gluons are soft. 
This term 
therefore does not
contribute to the differential 
three jet cross section, but to the ${\cal O}(\alpha_s^2)$ 
two parton cross section.
(Also the second
and the third term contain such contributions and the negative 
sign of the fourth term compensates the double counting.) 
\par
We will now derive the complete contribution from soft gluons
to the three jet cross section at NLO. 
We start by discussing the contribution from $|S_1|^2$. 
With our choice
of the Heaviside functions defining the soft region, it is simple to
analytically integrate the eikonal factors (\ref{eikonal})
in d dimensions over the soft gluon momentum. Let us consider
the case where $k_3\to 0$ (second term in  (\ref{S1decomp})). 
The case $k_4\to 0$ can
be treated in complete analogy. Since the two gluons
are identical particles, the overall soft factor
multiplying the Born cross section for $e^+e^-\to Q\bar{Q}g$
is determined by considering the contributing from one soft gluon 
only and leaving out the identical particle factor $(1/2!)$.
In the c.m. system of the heavy
quark $Q(k_1)$  and the hard gluon $g(k_4)$, the eikonal factor
reads
\bea 
f_l(1,3,4) =  \frac{1}{E_3^2E_1}
\left(\frac{2\sqrt{t_{14}+m^2}}{(1+\cos\theta)
(1-\beta_{14}\cos\theta)}-\frac{m^2}{E_1(1-\beta_{14}\cos\theta)^2}
\right),
\eea  
where $E_{1,3}$ are the heavy quark and soft gluon energies
in that system,
$\theta$ is the angle between the heavy quark and the soft gluon,
and $\beta_{14}=t_{14}/(t_{14}+2m^2)$. In the same system we have
(with $d=4-2\epsilon$)
\bea
\overline{\Theta}_{134}\frac{d^{d-1}k_3}{(2\pi)^{d-1}2E_3}=
\Theta(\frac{s_{min}}{\sqrt{t_{14}+m^2}}-E_3)\frac{1}{8\pi^2}
\frac{(4\pi)^{\epsilon}}{\Gamma(1-\epsilon)}
E_3^{1-2\epsilon}(\sin\theta)^{-2\epsilon}dE_3d\cos\theta.
\eea
The integration over the soft gluon momentum $k_3$ can now
be carried out without difficulty.\par
The soft contribution from $|S_2|^2$ is obtained in the same manner by
switching the roles of the heavy quark $Q(k_1)$ and the heavy
antiquark $\bar{Q}(k_2)$. For the term subleading in color, we write
\bea \label{S12decomp}
|S_1+S_2|^2\ &=&\  \left(\Theta_{132}+\overline{\Theta}_{132}\right)
\left(\Theta_{142}+\overline{\Theta}_{142}\right)|S_1+S_2|^2 \nonumber \\
\ &=&\ \left(\Theta_{132}\Theta_{142} + \overline{\Theta}_{132}
+ \overline{\Theta}_{142} -  \overline{\Theta}_{132} \overline{\Theta}_{142}
\right)|S_1+S_2|^2.
\eea
Since the amplitude $S_1+S_2$ is a QED-like contribution
with massive quarks, it does not induce collinear
but only soft singularities \cite{MaSi75}. 
For $k_{3,4}\to 0$ we choose 
the quark-antiquark c.m. system to evaluate the eikonal factors.
The complete soft factor $S$ multiplying the
squared Born matrix element  $\spincolorsum
|{\cal T}^{\scriptsize{\mbox{Born}}}(e^+e^-\to Q\bar{Q}g)|^2$
which we obtain by adding all contributions reads, if we relabel
the remaining hard gluon momentum with $k_3$: 
\bea
\label{softfactor}
S\ &=&\ 
\frac{\alpha_s}{4\pi}\,N_C\,\,\frac{1}
    {\Gamma(1-\epsilon)}\,
    \left(\frac{4\pi\mu^2}{s_{min}}\right)^{\epsilon}
\Bigg[
    \Bigg\{
        \left(\frac{s_{min}}{t_{13}+m^2}\right)^{-\epsilon}
       \Bigg( \frac{1}{\epsilon^2}
        -  \frac{1}{\epsilon}
        \bigg[\ln\left(1+\frac{t_{13}}{m^2}\right)
 \nonumber \\ \ &+& 2\,\ln(2) - 1 \bigg]
     - \frac{\pi^2}{6}+2\ln^2(2)-2\ln(2)
        +\Big[2\ln (2)+\frac{2m^2}{t_{13}}+1\Big]\,
        \ln\left(1+\frac{t_{13}}{m^2}\right) \nonumber \\ \ &-& \
        \frac{1}{2}\ln^2\left(1+\frac{t_{13}}{m^2}\right)
      - 2\,\Li{\frac{t_{13}}{t_{13}+m^2}}
        \Bigg) + (t_{13}\leftrightarrow t_{23})
     \Bigg\} \nonumber \\ \ &-&\ \frac{1}{N_C^2}
    \left(\frac{s_{min}}{t_{12}+2m^2}\right)^{-\epsilon} \frac{1}{\beta}
    \Bigg(\frac{1}{\epsilon}\left[\,2\beta- (1+\beta^2)\ln(\omega)\right]
     - 4\beta\ln(2) + 2\ln(\omega)
    \nonumber \\
    \ &+&\ 2 \ln(2) (1+\beta^2)\ln(\omega)
    -\frac{1+\beta^2}{2}\ln^2(\omega)-
    2(1+\beta^2)\Li{\frac{2\beta}{1+\beta}} \Bigg)\Bigg]
    + {\cal O}(\epsilon) ,
\eea
\noindent where $\beta$ and
$\omega$ have been defined in terms of the scaled quark and
antiquark c.m. energies $x,\xb$ in
(\ref{beom}). (We have $t_{12}=s(1-x_g-2z)$,
$t_{13}=s(1-\xb)$, $t_{23}=s(1-x)$.)
\par
The leading color contribution $|S_1|^2+|S_2|^2$ 
to the matrix element
(\ref{color2}) also contains collinear singularities. We isolate them
by writing
\bea\label{coll1}
\Theta_{134}\Theta_{243}|S_1|^2 \ &=&\ \big[\Theta_{134}\Theta_{243}
-\Theta(t_{13}-2s_{min})\Theta(t_{24}-2s_{min})\Theta(s_{min}-t_{34})
\big]|S_1|^2\nonumber \\ \ &+&\ \Theta(t_{13}-2s_{min})
\Theta(t_{24}-2s_{min})\Theta(s_{min}-t_{34})|S_1|^2.
\eea
The first term in (\ref{coll1}) is now free of singularities over the
whole four parton phase space allowed by the  Heaviside 
functions, while the 
second term contains the contribution of two collinear gluons which are both
{{\it not}} soft. By construction we thus avoid an overlap of
the soft and the collinear part of phase space.
The resolved part of the 
differential
cross section for  $e^+e^-\to Q\bar{Q}gg$
entering (\ref{resolved}) may thus be defined as:
\bea 
\label{4resolved}
d\sigma_2^R(e^+e^-\to Q\bar{Q}gg) \ &=&\
\frac{1}{2s}\frac{1}{2!}
\left[\prod_{i=1}^4\frac{d^{3}k_i}{(2\pi)^{3}2E_i}\right](2\pi)^4
\delta(k-\sum_{i=1}^4 k_i)
\nonumber \\ \ &\times&\ \frac{N_C^2-1}{2}\frac{N_C}{2}
\spinsum\bigg\{- \frac{1}{N_C^2}\Theta_{132}\Theta_{142}
|S_1+S_2|^2 \nonumber \\ \ &+& \ 
\left[\Theta_{134}\Theta_{243}-
\Theta(t_{13}-2s_{min})\Theta(t_{24}-2s_{min})\Theta(s_{min}-t_{34})\right]
|S_1|^2 \nonumber \\ \ &+& \  \left[\Theta_{234}\Theta_{143}-
\Theta(t_{23}-2s_{min})\Theta(t_{14}-2s_{min})\Theta(s_{min}-t_{34})\right]
|S_2|^2. \bigg\}\nonumber \\ & & 
\eea
\par
The remaining calculation is completely analogous to the massless
case \cite{GiGl92}; we include it here for completeness.\par
In the limit $k_3\parallel k_4$ we define
\bea 
&& k_3
{\buildrel
k_3 \| k_4\over \longrightarrow}  \xi k_h, \nonumber \\
&& k_4 
{\buildrel k_3 \| k_4\over \longrightarrow}  (1-\xi)k_h,
\eea
with $k_h=k_3+k_4$. In this limit,
\bea\label{S1coll}
\frac{N_C^2-1}{2}\frac{N_C}{2}\spinsum|S_1|^2 \ \ &
{\buildrel
k_3\|k_4\over \longrightarrow}&\ \  
\frac{g_s^2N_C}{2}f^{gg\to g}
\spincolorsum|{\cal T}^{\scriptsize{\mbox{Born}}}(e^+e^-\to Q\bar{Q}g)|^2,
\eea
where
\bea
f^{gg\to g}=\frac{2}{t_{34}}\frac{1+\xi^4+(1-\xi)^4}{\xi(1-\xi)}
\eea
is proportional to the Altarelli-Parisi splitting function.
The collinear behavior of $|S_2|^2$ is identical to (\ref{S1coll}),
leading to an overall factorization of the squared matrix element
(\ref{color2}) in the collinear limit. Further, with
$t_{13}=\xi t_{1h},\ t_{24}= ( 1-\xi)t_{2h}$ in this limit,
\bea
\lefteqn{\Theta(t_{13}-2s_{min})\Theta(t_{24}-2s_{min})\Theta(s_{min}-t_{34})
\frac{d^{d-1}k_3}{(2\pi)^{d-1}2E_3}\frac{d^{d-1}k_4}{(2\pi)^{d-1}2E_4}
= }  \hspace{5cm}\nonumber \\ 
& &
\Theta(\xi t_{1h}-2s_{min})\Theta((1-\xi)t_{2h}-2s_{min})
\Theta(s_{min}-t_{34}) \nonumber \\ & &    
\frac{1}{16\pi^2}\frac{(4\pi)^{\epsilon}}{\Gamma(1-\epsilon)}
\left[t_{34}\xi(1-\xi)\right]^{-\epsilon}dt_{34}d\xi
\frac{d^{d-1}k_h}{(2\pi)^{d-1}2E_h}.
\eea

After integration over $\xi$ and $t_{34}$, summing the contributions
from $|S_1|^2$ and $|S_2|^2$, and relabelling
$k_h \to k_3$, we get for the collinear factor $C^{gg}$
multiplying  $\spincolorsum
|{\cal T}^{\scriptsize{\mbox{Born}}}(e^+e^-\to Q\bar{Q}g)|^2$
(a statistical factor 1/2! is included):
\bea
C^{gg}\ &=&\  
\frac{\alpha_s}{2\pi}\,N_C\,\,\frac{1}{\Gamma(1-\epsilon)}\,
    \left(\frac{4\pi\mu^2}{s_{min}}\right)^{\epsilon}
 \Bigg\{\frac{1}{\epsilon}\left[\ln(\xi_1)+\ln(\xi_2)+\frac{11}{6}\right]
\nonumber \\ \ &-& \ \frac{\pi^2}{3}+\frac{67}{18}-\frac{1}{2}\ln^2(\xi_1)
-\frac{1}{2}\ln^2(\xi_2)\Bigg\}+{\cal O}(\epsilon) ,
\eea
where $\xi_1=2s_{min}/t_{13},\ \xi_2=2s_{min}/t_{23}$.
\par

As mentioned before, the contribution of the process 
$e^+e^-\to Q(k_1)\bar{Q}(k_2)
q(k_3)\bar{q}(k_4)$ to
the three jet cross section is singular only when the two
massless quarks become collinear. To isolate the singular
term, we use 

\bea
|{\cal T}(e^+e^-\to Q\bar{Q}q\bar{q})|^2
= \left[\Theta(t_{34}-s_{min})+\Theta(s_{min}-t_{34})\right]
|{\cal T}(e^+e^-\to Q\bar{Q}q\bar{q})|^2.
\eea

For the resolved part of the cross section we may therefore
write
\bea \label{4QQqq}
d\sigma_2^R(e^+e^-\to Q\bar{Q}q\bar{q})  \ &=&\  \Theta(t_{34}-s_{min})
d\sigma_2(e^+e^-\to Q\bar{Q}q\bar{q}).
\eea

In the phase space region defined by $\Theta(s_{min}-t_{34})$,
we use the collinear limit

\bea
\spincolorsum|{\cal T}(e^+e^-\to Q\bar{Q}q\bar{q})|^2
\ {\buildrel k_3\|k_4\over \longrightarrow} 
\ \frac{g_s^2 n_f^{\rm ms}} {2} f^{q\bar{q}\to g}
\spincolorsum|{\cal T}^{\scriptsize{\mbox{Born}}}(e^+e^-\to Q\bar{Q}g)|^2,
\eea
where
\bea
f^{q\bar{q}\to g}=\frac{2}{t_{34}}\frac{\xi^2+(1-\xi)^2-\epsilon}
{1-\epsilon},
\eea
and $n_f^{\rm ms}$ is the number of massless flavors.
After integration over the collinear phase space we get for
the collinear factor $C^{q\bar{q}}$ multiplying the squared Born
matrix element for $e^+e^-\to Q\bar{Q}g$:
\bea
C^{q\bar{q}}=\frac{\alpha_s}{2\pi}n_f^{\rm ms}\frac{1}{\Gamma(1-
\epsilon)}\left(\frac{4\pi\mu^2}{s_{min}}\right)^{\epsilon}
\left\{-\frac{1}{3\epsilon}-\frac{5}{9}\right\} +
{\cal O}(\epsilon).
\eea
\par

We have thus derived  the 
following differential
cross sections for three resolved partons 
entering (\ref{resolved}):
\bea 
\label{3resolved}
d\sigma_2^R(e^+e^-\to Q\bar{Q}g) \ &=&\ 
\frac{1}{2s}
\left[\prod_{i=1}^3\frac{d^{3}k_i}{(2\pi)^{3}2E_i}\right](2\pi)^4
\delta(k-\sum_{i=1}^3 k_i)\Theta_{132} \nonumber \\ &\times&
\lim_{\epsilon\to 0}
\bigg\{{\cal T}^{\scriptsize{\mbox{Born}}}
(e^+e^-\to Q\bar{Q}g){\cal T}^{\scriptsize{\mbox{virtual}}}
(e^+e^-\to Q\bar{Q}g)^{\ast}+{\mbox{h.c.}} \nonumber \\
\ &+& \ \bigg[S+C^{gg}
+C^{q\bar{q}}\bigg]
|{\cal T}^{\scriptsize{\mbox{Born}}}
(e^+e^-\to Q\bar{Q}g)|^2\bigg\}.
\eea
\par
In particular, we have for the function $F_1$ defined in
(\ref{parton1}), (\ref{parton2}),
 which
determines double differential
cross section for three resolved partons: 
\bea
F_1 = F_1^{(0)}+F_1^{(1)}+ O(\alpha_s^2),
\eea
with $F_1^{(0)}$ given in (\ref{Fi01}), (\ref{Fi02}), and
\bea
F_1^{(1)} = \lim_{\epsilon\to 0}
\bigg\{F_1^{\scriptsize{\mbox{virtual}}}+
F_1^{\scriptsize{\mbox{soft+collinear}}}\bigg\},
\eea
where $F_1^{\scriptsize{\mbox{virtual}}}$ has been listed
in (\ref{F_1v})--(\ref{counterext}) and
\bea
F_1^{\scriptsize{\mbox{soft+collinear}}}=
 \bigg[S+C^{gg}+C^{q\bar{q}}\bigg]\bigg\{F_1^{(0)}
-\epsilon\bigg[g_{PC}^{VV}\frac{x_g^2B}{2}
+g_{PC}^{AA}\frac{x_g^2B(1+2z)}{2}\bigg]\bigg\}.
\eea
One can easily verify
that $F_1^{(1)}$ is finite (the poles in 
$F_1^{\scriptsize{\mbox{virtual}}}$ and 
$F_1^{\scriptsize{\mbox{soft+collinear}}}$ exactly cancel).
The dependence of $F_1^{(1)}$ on $s_{min}$ cancels 
in its contribution to an observable quantity (like the three jet
cross section) against 
the $s_{min}$ dependence
from the contributions of four resolved partons in the
limit $s_{min}\to 0$. The latter are
obtained numerically starting from expressions (\ref{4bs}),
(\ref{4resolved}), and (\ref{4QQqq}).
Examples will be given in the next section.
We have derived analytic expressions also for all the
other functions $F_i^{(1)},\ i=2,\ldots,9$ 
which enter the fully differential 
cross section (\ref{parton1}) for three resolved partons at NLO;
but we will discuss them elsewhere \cite{future}. 
\end{section}
\begin{section}{Numerical results}
\label{numerics}
\setcounter{equation}{0}
In this section we show results for  
some observables involving massive quarks.
We carry out the necessary numerical integrations of our matrix
elements with the help of {\sc Vegas} \cite{Vegas}. All quantities
are calculated by expanding in $\alpha_s$ to NLO accuracy.
\par
The three jet cross section for $b$ quarks 
as a function of the
jet resolution parameter $y_{cut}$ in the JADE and Durham scheme 
as well as an observable sensitive to the mass of the $b$ quark at the $Z$ pole
have already been presented in \cite{BeBrUw97}.
Here we start our discussion by demonstrating the independence
of physical quantities on the parameter $s_{min}$ as $s_{min}\to 0$. We choose
as an example the three jet fraction for $b$ quarks,
\bea \label{f3b}
f_3^b(y_{cut})=\frac{\sigma_3^b(y_{cut})}{\sigma_{tot}^b}.
\eea
In (\ref{f3b}), the numerator $\sigma_3^b$ 
is defined as the three jet cross section
for events in which at least two jets containing a $b$ or $\bar{b}$ quark
remain after the clustering procedure. This requirement  ensures that
the cross section stays finite also in the limit $m_b\to 0$. 
The contribution of the process
$e^+e^-\to Z,\gamma^{*}\to q\bar{q}g^{*}\to q\bar{q}b\bar{b}$ to the three
jet cross section with {\it one} tagged $b$ quark develops 
large logarithms $\ln(m_b^2)$ 
-- which find no counterpart in the virtual corrections
against which they can cancel --
when the $b\bar{b}$ pair is clustered into a single jet. 
In principle, there are two distinct possibilities to handle this problem:
One may either impose suitable experimental requirements/cuts to get rid
of events with two light quark jets and one jet containing a $b\bar{b}$
pair (the definition for $\sigma_3^b$
chosen by us is an example for this), or one can 
improve the fixed order calculation by absorbing the large logarithm
into a fragmentation function for a gluon into a heavy quark.
A detailed discussion of this issue will be presented elsewhere 
\cite{BeBrUw297}. Note that $\sigma_{tot}^b$ has to be calculated only
to order $\alpha_s$ for the NLO prediction of $f_3^b$; hence the
prescription to handle the $g^{\ast}\to b\bar{b}$ contribution
does not affect the evaluation of the denominator of
(\ref{f3b}) at this order.
\par
Figs. \ref{fig:sminja} and  \ref{fig:smindu} show the three jet 
fraction $f_3^b$ in the JADE 
and Durham  scheme at NLO as a function
of $y_{min}=s_{min}/(sy_{cut})$ at a fixed  value of $y_{cut}=0.03$ 
and $\sqrt{s}=m_Z=91.187$ GeV. The error bars are due to the 
numerical integration.
For the renormalization scale we take in these
plots $\mu=\sqrt{s}$. 
As to the mass parameter, we use 
$m^{\scriptsize{\mbox{$\msbar$}}}_b(\mu)$ defined
in the $\msbar$ scheme at the scale $\mu$. The asymptotic
freedom property of QCD predicts that this mass 
parameter decreases when being evaluated at a higher
scale. (A number of low energy determinations of the $b$ quark mass have
been made; see for instance \cite{Na94,Ne94,Da94,Ja97} and references therein.)
With $m^{\scriptsize{\mbox{$\msbar$}}}_b(\mu=m_b)=4.36$ GeV \cite{Ne94}
and $\alpha_s(m_Z)=0.118$ \cite{PDG} as an input and employing 
the standard renormalization group evolution of the 
coupling and the quark masses, we use the value 
$m^{\scriptsize{\mbox{$\msbar$}}}_b(\mu=m_Z)=3$ GeV.
One clearly sees that $f_3^b$ reaches a plateau
for small values of $y_{min}$. The error in the numerical integration
becomes bigger as $y_{min}\to 0$.  In order to keep this error as small as 
possible without introducing a systematic error from using the soft and collinear
approximations, we take in the following $y_{min}=10^{-2}$ for the JADE algorithm and 
$y_{min}=0.5\times 10^{-2}$ for the Durham algorithm. At these values, 
the dominant $s_{min}$-dependent individual contributions from three and 
four resolved partons are about
a factor of 2.5 (JADE) and 4 (Durham) larger than the sum. 
\par 
In Figs. \ref{fig:ycutja} and  \ref{fig:ycutdu} 
we plot  $f_3^b$ as a function of 
$y_{cut}$ at LO and NLO, again at $\sqrt{s}=m_Z$.
 The QCD corrections to the LO result are quite 
sizable as known also in the massless case.
The renormalization scale dependence (where $\mu$ is varied
between $m_Z/2$ and $2m_Z$), which is also shown in
Figs. \ref{fig:ycutja} and \ref{fig:ycutdu}, is modest in the whole $y_{cut}$ 
range exhibited for the Durham
and above $y_{cut}\sim 0.01$ for the JADE
algorithm. Below this value perturbation theory does not yield
reliable results  in the JADE scheme.
In Figs. \ref{fig:renormja} and \ref{fig:renormdu} we take a closer look
on the scale dependence of $f_3^b$, now using the on-shell mass 
renormalization scheme. We vary the scale $\mu$ between $m_Z/16$ and
$2m_Z$ for a fixed value $y_{cut}=0.2\times 10^{-3/10}\approx 0.1$ and on-shell masses
$m^{\rm pole}_b=3$ GeV and $m^{\rm pole}_b=5$ GeV. In Fig. 
\ref{fig:renormja} we see that the scale dependence of
the LO result (which is solely due to
the scale dependence of $\alpha_s$ at this order) 
in the JADE algorithm  amounts
to about 100\% in the $\mu$ interval shown. The 
inclusion of the $\alpha_s^2$ corrections reduces the scale dependence
significantly; the NLO result for $f_3^b$ at $\mu=2m_Z$ is about 30\% smaller
than the NLO result at $\mu=m_Z/16$. 
In the case of the Durham algorithm, the difference between 
$f_3^b$ at $\mu=2m_Z$ and at $\mu=m_Z/16$ is reduced from
about 100\% at LO to about 10\% at NLO.         
\par
The effect of the $b$ quark mass may be illustrated by
looking at the double ratio 
\bea
{\cal C}(y_{cut})=\frac{f_3^b(y_{cut})}{f_3^{{\rm incl.}}(y_{cut})},
\eea
where the denominator is the three jet fraction when summing
over all active quark flavors, which is given to a very good 
approximation by the massless NLO result \cite{ElRoTe81}-
\cite{KuNa89}. Similar double ratios have been studied
in \cite{BeBrUw97} and \cite{BiRoSa95}, \cite{Ro96}, \cite{BiRoSa97}. 
In Fig. 
\ref{fig:edep}  we
plot ${\cal C}$ as a function of the c.m. energy at $y_{cut}=0.08$
for the JADE algorithm. The running of $\alpha_s$  
is taken into account in the
curves, where we again use as an input $\alpha_s(\mu=m_Z)=0.118$.
For the dashed (LO) and full (NLO) curve we use the running 
mass $m^{\scriptsize{\mbox{$\msbar$}}}_b$ with
$m^{\scriptsize{\mbox{$\msbar$}}}_b(\mu=m_Z)=3$ GeV. 
For all energies, the
renormalization scale is set to $\mu=\sqrt{s}$. For comparison we
also show the LO result for a fixed value of the $b$ quark mass 
$m_b=
4.7$ GeV (dash-dotted curve), which is the corresponding
value of the pole mass. One clearly sees that the effect 
of the $b$ quark mass gets larger for smaller c.m. energies.
\par
Another interesting quantity to study mass effects is the 
differential two jet rate \cite{Op90} defined as
\bea
D_2(y)=\frac{f_2(y)-f_2(y-\Delta y)}{\Delta y},
\eea
where $f_2(y)$ is the two jet fraction at $y=y_{cut}$ for
a given jet algorithm. The advantage of $D_2$ over the
three jet fraction $f_3$ lies in the fact that the statistical
errors in bins of $D_2(y)$ are independent from each other
since each event enters the distribution only once. To order
$\alpha_s^2$, $D_2(y)$ can be calculated from the three- and
four jet fractions using the identity 
\bea
1=f_2+f_3+f_4+O(\alpha_s^3).
\eea
We define
\bea
{\cal D}(y)=\frac{D_2^b(y)}{D_2^{\rm incl}(y)},
\eea
where we -- as in the case of the quantity ${\cal C}$ -- 
use the massless NLO result to evaluate
the denominator. We plot our results for ${\cal D}(y)$
in Figs. \ref{fig:d2ja} and \ref{fig:d2du}, again
for $\sqrt{s}=\mu=m_Z$. The full circles show the
NLO results for
$m^{\scriptsize{\mbox{$\msbar$}}}_b(\mu=m_Z)=3$ GeV.
For the $O(\alpha_s^2)$ 
contribution of  $f_3^b(y_{cut})$ to ${\cal D}$ 
we use a fit to the numerical results.  For
comparison, the
squares (triangles) are the LO results for
$m_b=3$ GeV ($m_b=5$ GeV). The horizontal bars
show the size of the bins in $y_{cut}$.
The effects of the $b$ quark mass are of the order
of 5\% or larger at small values of $y_{cut}$.
\par
Finally we discuss an event shape variable which does not depend 
on the jet finding algorithm, namely thrust \cite{Fa77}, defined
as the sum of the lengths of the longitudinal momenta of the final 
state particles relative to the axis ${\bf n}$ chosen to
maximize this sum,
\bea
T= {\mbox{max\ }} 
\frac{\sum_i|{\bf k}_i\cdot {\bf n}|}{\sum_i|{\bf k}_i|}.
\eea
We may write the thrust distribution for $b$ quarks as
\bea
\label{thrdist}
\frac{1}{\sigma_{\rm tot}^b}\frac{d\sigma^b}{dT} = 
\frac{\alpha_s}{2\pi}c_1+\left(\frac{\alpha_s}{2\pi}\right)^2c_2+O\left(\alpha_s^3\right).
\eea
In the massless calculation at $\mu=\sqrt{s}$, the coefficients
$c_1$ and $c_2$ are independent of the c.m. energy. This
changes in the massive case as shown in Figs. 
 \ref{fig:thrLO} and \ref{fig:thrNLO}. For these and the 
following plots of the thrust distribution we exclude
the singular two-jet region near $T=1$. We use
a slicing parameter $s_{min}=1{\mbox{\ GeV}}^2$. 
We now only require the tagging of {\it one} $b$ quark and 
omit the contributions from the process
$e^+e^-\to q\bar{q}g^\ast\to q\bar{q}b\bar{b}$.
In Fig. \ref{fig:thr30}  we plot the thrust distribution (\ref{thrdist})
at $\sqrt{s}=\mu=30$ GeV with $\alpha_s(\mu=\sqrt{s})=0.142$ and
 $m^{\scriptsize{\mbox{$\msbar$}}}_b(\mu=\sqrt{s})=3.33$ GeV.
Shown separately (and not included in the NLO histogram) 
is the contribution from  
$e^+e^-\to q\bar{q}g^{\ast}\to q\bar{q}b\bar{b}$ calculated ``naively'',
i.e. directly from the matrix element without imposing cuts.
The LO and NLO thrust distributions for $b$ quarks at 
$\sqrt{s}=m_Z$ depicted in Fig. \ref{fig:thr91} 
are almost identical to the corresponding 
distributions at $\sqrt{s}=30$ GeV. The reason is that the change
in the coefficients $c_{1,2}$ due to mass effects as exhibited 
in Figs. \ref{fig:thrLO} and \ref{fig:thrNLO} is almost 
exactly compensated by the larger value of $\alpha_s$ at
$\mu=30$ GeV. The contribution from $g^{\ast}\to b\bar{b}$
splitting is bigger at the higher energy scale.

\end{section}
\begin{section}{Summary}
  \label{concl}
In view of the large number of jet events collected both at LEP
and SLC it is desirable for precision tests of 
QCD to use NLO partonic matrix elements
that include the full quark mass dependence. We have presented in
this article the necessary ingredients to calculate any jet quantity
that gets contributions from three- and four jet final states
involving massive quarks at order $\alpha_s^2$.
In particular, we have derived an explicit analytic expression
for the virtual corrections to $e^+e^-\to Q\bar{Q}g$.
Our approach involves a modification of the phase space slicing
method incorporating massive quarks. 
The independence of physical quantities on the slicing
parameter $s_{min}$ is a crucial test of the overall
consistency and has been carefully checked. 
As to numerical results
we have evaluated the three jet fraction for $b$ quarks 
and compared it to the inclusive
three jet fraction evaluated for massless quarks 
at different c.m. energies. 
We have further studied the dependence
of our results on the renormalization scale. If we take this dependence
as an estimate of the neglected higher order corrections, 
we may conclude that they are of moderate size,
in particular in the case of jet rates computed using the Durham
algorithm.
A particularly interesting quantity to study mass effects is the
differential two jet rate. At the $Z$ pole and for small $y_{cut}$,
the effects of the $b$ quark mass are of the order of 5\% or larger.
As a final example we studied the thrust distribution for $b$ quark samples 
at $\sqrt{s}=m_Z$ and $\sqrt{s}=30$ GeV. 
Future work will include the calculation of several other three jet
quantities, in particular also parity violating observables.
\end{section}
\section*{Acknowledgments}
We would like to thank M. Flesch and P. Haberl for discussions,
and especially W. Bernreuther for the
collaboration on this project and for his
careful reading of the manuscript.

\appendix
\section{Appendix}
\setcounter{equation}{0}
In (\ref{F_1v2}), we defined the 
finite contributions $F_1^{{\rm lc},VV(AA)}$ and 
$F_1^{{\rm sc},VV(AA)}$ to the function 
$F_1^{{\rm virtual},VV(AA)}$, which are 
generated by  the interference of the loop diagrams
Figs. 2(a)-(k) with the Born graphs. In this appendix, 
we give explicit expressions for these contributions
in terms of finite parts of one-loop integrals times
coefficient functions.
For the two- and three-point loop
integrals we use the notation of \cite{PaVe79}, i.e. all
integrals are labelled with respect to the following
box integrals:
\bea
  D_0&=&\frac{1}{i\pi^2}
  \int\!\!\frac{\left(2\pi\mu\right)^{-2\epsilon}d^{4-2\epsilon}l}
{(l^2+i\varepsilon)((l+k_3)^2+i\varepsilon)
    ((l+k_{13})^2-m^2+i\varepsilon)((l-k_2)^2-m^2+i\varepsilon)}, \nonumber \\
  D_0^{{\rm sc},1}&=&\frac{1}
{i\pi^2}\int\!\!
  \frac{\left(2\pi\mu\right)^{-2\epsilon}d^{4-2\epsilon}l}{(l^2+i\varepsilon)((l+k_1)^2-m^2+i\varepsilon)
    ((l+k_{13})^2-m^2+i\varepsilon)((l-k_2)^2-m^2+i\varepsilon)}, \nonumber \\
D_0^{{\rm sc},2}&=&D_0^{{\rm sc},1}\big|_{k_1\leftrightarrow k_2},
\eea
with $k_1^2=k_2^2=m^2$, $k_3^2=0$, and $k_{13}=k_1+k_3$. In the following, 
$C_0(1,2,3)$ denotes the integral obtained from $D_0$ by omitting the
fourth denominator, etc. The nonvanishing one-point function
is defined by $A(zs)=(2\pi\mu)^{-2\epsilon}\int\!\!d^dl[l^2-m^2+i\varepsilon]^{-1}$.
As stated in section \ref{virtual},
we eliminate the above scalar box integrals in $d=4-2\epsilon$ dimensions
in favor of the box integrals in  $d=6-2\epsilon$ dimensions.
\par
We write 
\bea \label{lcdecomp}
F_1^{{\rm lc},VV(AA)} &=&
d_{d=6}^{VV(AA)}{\mbox{Re}}\,D_0^{d=6}
\nonumber \\ &+&
c_{123}^{VV(AA)}{\mbox{Re}}\,\overline{C}_0(1,2,3)+
c_{124}^{VV(AA)}{\mbox{Re}}\,\overline{C}_0(1,2,4)
\nonumber \\
&+& 
c_{134}^{{\rm lc},VV(AA)}{\mbox{Re}}\,C_0(1,3,4)+
c_{234}^{{\rm lc},VV(AA)}{\mbox{Re}}\,C_0(2,3,4)
\nonumber \\
&+& 
b_{13}^{{\rm lc},VV(AA)}{\mbox{Re}}\,\overline{B}_0(1,3)+
b_{24}^{{\rm lc},VV(AA)}{\mbox{Re}}\,\overline{B}_0(2,4)
\nonumber \\ &+&
b_{14}^{{\rm lc},VV(AA)}{\mbox{Re}}\,\overline{B}_0(1,4)
+
b_{34}^{{\rm lc},VV(AA)}{\mbox{Re}}\,\overline{B}_0(3,4)
\nonumber \\ &+&
a^{{\rm lc},VV(AA)}{\mbox{Re}}\,\overline{A}+
k^{{\rm lc},VV(AA)}.
\eea

\bea \label{scdecomp}
F_1^{{\rm sc},VV(AA)} &=&
\tilde{d}_{d=6}^{VV(AA)}{\mbox{Re}}\,D_0^{{\rm sc},1,d=6}
+\tilde{d'}_{d=6}^{VV(AA)}{\mbox{Re}}\,D_0^{{\rm sc},2,d=6}
\nonumber \\ &+&
\tilde{c}_{123}^{VV(AA)}{\mbox{Re}}\,C_0^{{\rm sc},1}(1,2,3)+
\tilde{c}_{134}^{VV(AA)}{\mbox{Re}}\,C_0^{{\rm sc},2}(1,3,4)
\nonumber \\
&+& 
\tilde{c}_{124}^{VV(AA)}{\mbox{Re}}\,
\overline{C}_0^{{\rm sc},1}(1,2,4)
+
c_{134}^{{\rm sc},VV(AA)}{\mbox{Re}}\,C_0(1,3,4)
\nonumber \\
&+& 
c_{234}^{{\rm sc},VV(AA)}{\mbox{Re}}\,C_0(2,3,4)+
\tilde{c}_{234}^{VV(AA)}{\mbox{Re}}\,C_0^{{\rm sc},1}(2,3,4)
\nonumber \\ &+&
b_{13}^{{\rm sc},VV(AA)}{\mbox{Re}}\,\overline{B}_0(1,3)+
b_{24}^{{\rm sc},VV(AA)}{\mbox{Re}}\,\overline{B}_0(2,4)
\nonumber \\ &+&
b_{14}^{{\rm sc},VV(AA)}{\mbox{Re}}\,\overline{B}_0(1,4)
+
b_{34}^{{\rm sc},VV(AA)}{\mbox{Re}}\,\overline{B}_0(3,4)
\nonumber \\ &+&
\tilde{b}_{24}^{VV(AA)}{\mbox{Re}}\,\overline{B}^{{\rm sc},1}_0(2,4)+
a^{{\rm sc},VV(AA)}{\mbox{Re}}\,\overline{A}+
k^{{\rm sc},VV(AA)}.
\eea

The symbol 
$\overline{I}$ means that the {\it finite}
part of the UV or IR divergent integral $I$ is to be taken.
\par
The real parts of the integrals appearing in (\ref{lcdecomp})
and (\ref{scdecomp}) read (recall that $z=m^2/s,\ x_g=2-x-\xb$):

\begin{eqnarray}
%
%
%
%
 {\mbox{Re}}\,D_0^{d=6}&=& \frac{\pi}{sx_g}\Bigg\{
  \frac{(1-\xb)\alpha}{(\alpha-x_+^s)(\alpha-x_-^s)}
  \left[-\frac{\pi^2}{6}+\Li{\frac{\alpha-1}{\alpha}}\right]\nonumber\\
  &+&\frac{(1-\xb)x_+^s}{(\alpha-x_+^s)(x_+^s-x_-^s)}\Bigg[\,
  \frac{\pi^2}{6} 
  - \Li{\frac{x_+^s-1}{x_+^s}} 
  + \Li{\frac{x_+^s-1}{x_+^s-\rho}}\nonumber\\
  &+&\Li{\frac{x_+^s-\rho}{x_+^s}}
  - \Li{\frac{1-\rho-x_+^s}{1-x_+^s}}
  - \Li{\frac{-x_+^s}{1-x_+^s-\rho}}\nonumber\\
  &+&\frac{1}{2}\ln^2\left(\frac{x_+^s}{x_+^s-\rho}\right)
  -\frac{1}{2}\ln^2\left(\frac{1-x_+^s}{1-\rho-x_+^s}\right)
  \,\Bigg]\nonumber\\
  &+&\frac{(1-\xb)x_-^s}{(\alpha-x_-^s)(x_-^s-x_+^s)}\Bigg[\,
  \frac{\pi^2}{6} 
  - \Li{\frac{x_-^s-1}{x_-^s}} 
  + \Li{\frac{x_-^s-1}{x_-^s-\rho}}\nonumber\\
  &+&\Li{\frac{x_-^s-\rho}{x_-^s}}
  - \Li{\frac{1-\rho-x_-^s}{1-x_-^s}}
  - \Li{\frac{-x_-^s}{1-x_-^s-\rho}}\nonumber\\
  &+&\frac{1}{2}\ln^2\left(\frac{x_-^s}{x_-^s-\rho}\right)
  -\frac{1}{2}\ln^2\left(\frac{1-x_-^s}{1-\rho-x_-^s}\right)
  \,\Bigg]\Bigg\}\nonumber\\
  &+&(x \leftrightarrow \xb),  \\
%
%
  \Re \,C_0(1,2,3)&=& 
  \frac{1}{2s}\frac{1}{1-\xb}\Bigg\{\frac{1}{\epsilon^2}
  -\frac{1}{\epsilon}
  \left[\gamma+\ln\left(\frac{sz(\zeta-1)^2}{4\pi\mu^2}\right)\right]
  \nonumber\\
  &-&
  \frac{1}{12}\Bigg(15\pi^2+24\,\Li{\frac{1}{\zeta}}+12\ln^2(\zeta)\nonumber\\
 &-&6\left[\gamma
      +\ln\left(\frac{sz(\zeta-1)^2}{4\pi\mu^2}\right)\right]^2\Bigg)\Bigg\}, \\
%
%
\Re \,C_0(1,2,4)&=& \Re \,C_0(1,2,3)\Big|_{x\leftrightarrow\xb},
 \\
%
%
 \Re \,C_0(1,3,4) &=&-\frac{1}{s}\frac{1}{x_+^s-x_-^s}\,\Bigg\{
  -\frac{1}{2}\ln^2\left(\frac{x_+^s}{x_+^s-\rho}\right)
  +\frac{1}{2}\ln^2\left(\frac{x_+^s-1}{x_+^s+\rho-1}\right)\nonumber\\
  &-&\Li{\frac{x_+^s-1}{x_+^s-\rho}}
  +\Li{\frac{x_+^s}{x_+^s+\rho-1}}
  +\Li{\frac{x_+^s-1}{x_+^s}}\nonumber\\
  &-&\Li{\frac{x_+^s-\rho}{x_+^s}}
  +\Li{\frac{x_+^s+\rho-1}{x_+^s-1}}
  \Bigg\}
  + (x_+^s \leftrightarrow x_-^s), \\
%
%
  \Re \,C_0(2,3,4)&=& \Re \,C_0(1,3,4)\Big|_{x\leftrightarrow\xb}, \\
%
%
  \Re \,B_0(1,3)&=&\frac{1}{\epsilon}-\gamma
  -\ln\left(\frac{z\zeta s}{4\pi\mu^2}\right)
  +\frac{\zeta-1}{\zeta}\ln\left(\frac{\zeta}{\zeta-1}\right)
  +\frac{1}{\zeta}\ln(\zeta) + 2 , \\
%
%
  \Re \,B_0(2,4)&=& \Re \,B_0(1,3)\Big|_{x\leftrightarrow \xb},
 \\
%
%
   \Re \,B_0(1,4)&=&\frac{1}{\epsilon}-\gamma 
  - \ln\left(\frac{zs}{4\pi\mu^2}\right)+2, \\
%
%
  \Re \,B_0(3,4)&=&\frac{1}{\epsilon}-\gamma
  -\ln\left(\frac{zs}{4\pi\mu^2}\right)+2
  +(1-2\rho)\ln\left(\frac{\rho}{1-\rho}\right), \\
%
%
\Re \,A(zs)&=&zs\left(\frac{1}{\epsilon}-\gamma-\ln\left(\frac{zs}{4\pi\mu^2}
    \right) +1 \right), \\
%
%
%
%
  \Re\, C_0^{{\rm sc},1}(1,2,4)&=&\frac{1}{s}\,\frac{1}{1-x_g}\,\,\frac{1}{1-2\eta}
  \Bigg\{
  \,\frac{1}{\epsilon}\,\ln\left(\frac{\eta}{1-\eta}\right)\nonumber\\
  &-&2\,\Li{\frac{\eta}{1-\eta}}
  -\frac{1}{2}\ln^2\left(\frac{\eta}{1-\eta}\right)
  -\frac{2}{3}\pi^2\nonumber\\
  &-&\ln\left(\frac{\eta}{1-\eta}\right)\left[\gamma
    +\ln\left(\frac{(1-x_g)s}{4\pi\mu^2}\right)+2\ln(1-2\eta)\right]
  \Bigg\},  \\
%
%
 \Re\,  C_0^{{\rm sc},1}(1,2,3)&=&\frac{1}{s}\,\frac{1}{1-\xb}
  \left\{ \Li{\frac{1}{\zeta}}+\frac{1}{2}\ln^2(\zeta)-\frac{\pi^2}{6}\right\},\nonumber \\
  \Re\, C_0^{{\rm sc},2}(1,3,4)&=& 
\Re\,  C_0^{{\rm sc},1}(1,2,3)\Big|_{x\leftrightarrow \xb},
 \\
%
%
 \Re\, C_0^{{\rm sc},1}(2,3,4)&=&\frac{1}{sx_g}\Bigg\{
  \frac{1}{2}\ln^2\left(\frac{\rho}{1-\rho}\right)
  -\frac{1}{2}\ln^2\left(\frac{\eta}{1-\eta}\right)
\Bigg\}, \\
%
%
{\mbox{Re}}\,D_0^{{\rm sc},1,d=6}&=&
\frac{\pi}{2[(1-x_g)(1-x)(1-\xb)-zx_g^2]}\nonumber \\
&\times&
\bigg\{(1-\xb)(1-x_g)(1-x_g-4z) \nonumber \\
&\times&[-s(1-\xb)\Re\, D_0^{{\rm sc},1,d=4-2\epsilon}
+\Re\, C_0^{{\rm sc},1}(1,2,4)]\nonumber \\
&+&[(1-\xb)(1-x_g)-2zx_g]\nonumber \\
&\times&[(1-\xb)
\Re\, C_0^{{\rm sc},1}(1,2,3)-x_g
\Re\, C_0^{{\rm sc},1}(2,3,4)] \nonumber \\
&+&(1-\xb)[-\xb(1-x_g)+2z(x+\xb)]\Re\, C_0(1,3,4)
\bigg\}, \\
{\mbox{Re}}\, D_0^{{\rm sc},2,d=6}&=&
{\mbox{Re}}\, D_0^{{\rm sc},1,d=6}
  \Big|_{x\leftrightarrow\xb}, \\
%
%
%
   \Re\, B_0^{{\rm sc},1}(2,4)&=&\frac{1}{\epsilon}-\gamma
  -\ln\left(\frac{zs}{4\pi\mu^2}\right)+2
  +(1-2\eta)\ln\left(\frac{\eta}{1-\eta}\right),
\end{eqnarray}
where
\begin{eqnarray}
  \alpha &=& \frac{1-x}{x_g},\\ 
  \zeta &=& \frac{1-\xb +z}{z},\\ 
  \rho &=& \frac{1}{2} ( 1 - \sqrt{1-4z}),\\ 
  x_{\pm}^s &=& \frac{1}{2}(\xb\pm\sqrt{\xb^2-4z}),\\
  x_\mp^t&=&1-\frac{1}{2}(x\pm\sqrt{x^2-4z}),\\
  \eta &=& \frac{1}{2}\left(1-\sqrt{1-\frac{4z}{1-x_g}}\right),
\end{eqnarray}
and (cf. \cite{BeDe90})
\begin{eqnarray}
  \Re\, D_0^{{\rm sc},1,d=4-2\epsilon}&=&\frac{1}{s^2}\,\frac{1}{1-\xb}\,
  \frac{1}{1-x_g}\,\frac{1}{1-2\eta}\nonumber\\
  &\times&\Bigg\{
  \left[\frac{1}{\epsilon}-\gamma
    -\ln\left(\frac{sz(\zeta-1)^2}{4\pi\mu^2}\right)\right]
  \ln\left(\frac{\eta}{1-\eta}\right) - \pi^2 \nonumber\\
  &-& 2\,\Li{1-\frac{\eta}{1-\eta}\frac{\rho}{1-\rho}}
  - 2\,\Li{1-\frac{\eta}{1-\eta}\frac{1-\rho}{\rho}}\nonumber\\
  &+& \Li{1-\frac{\eta^2}{(1-\eta)^2}} 
  - \ln^2\left(\frac{\rho}{1-\rho}\right)
  \Bigg\}.
\end{eqnarray}

The coefficient functions read (where $B=1/(1-x)/(1-\xb)$,
and $\beta$ and $\omega$ have been defined in (\ref{beom})): 

\bea
d_{d=6}^{VV} \ &=&\ \frac{sB[1-x_g-zx_g^2B]}
{\pi}\bigg\{x^2 \nonumber \\ \ &+&\ 
4zB\left[-2x^2+5x+x\xb(x-2)-2-
z(2x^2+3(1-2x)+x\xb)\right]+({x\leftrightarrow \xb})
\bigg\},\nonumber \\ 
c_{123}^{VV} \ &=&\  -\frac{s}{1-x}\bigg\{x^2
+2zB[-3x^2+8x+2x\xb(x-2)-3-
zx_g^2]+({x\leftrightarrow \xb})
\bigg\}, \nonumber \\
c_{124}^{VV} \ &=&\ c_{123}^{VV}\Big|_{x\leftrightarrow \xb},
\nonumber \\
c_{134}^{{\rm lc},VV} \ &=&\ -\frac{zs}{1-x}
\bigg\{\frac{-4z(1+2z)}{1-\xb}-
\frac{2x^2+2x\xb-12x+9-4\xb-\xb^2}{1-x}
\nonumber \\ 
\ &-&\ 
\frac{z[13x^2-34x-12\xb+8x\xb+25-8z(2x+\xb-3)]}{(1-x)^2}
\nonumber \\ &+&
\frac{1}{\xb^2-4z}[-x^2-x^2\xb+8x-6x\xb+9\xb-7
\nonumber \\ &+& 
z(36x-x^2-60+29\xb-10x\xb)+8z^2(x+\xb-6)]
\nonumber \\
\ &-&\
\frac{3}{(\xb^2-4z)^2}
[-(1-x)^2\xb+z(8x^2-20x-16\xb+20x\xb
-5x^2\xb+12)\nonumber \\
\ &+&\ 4z^2(x^2-10x-5\xb+2x\xb+12)+16z^3]\bigg\}
, \nonumber \\
c_{234}^{{\rm lc},VV} \ &=&\ c_{134}^{{\rm lc},VV}\Big|_{x\leftrightarrow \xb},
\nonumber \\ 
b_{13}^{{\rm lc},VV} \ &=&\ \frac{3z}{1-x}-\frac{z(1-x)}{2(1-\xb)^2}
+\frac{-1+2z(5-2x)+4z^2(x-4)+16z^3}{2(1-\xb)(1-4z)}
\nonumber \\ 
\ &-&\ \frac{-1+x+2z(1-x)+z^2(x-2)+2z^3}
{2(1-\xb+z)(1-z)(1-x)}
+\frac{1}{2(1-x)(1-z)(1-4z)(\xb^2-4z)}
\nonumber \\ \ &\times&\ 
[-(1-x)(1-x_g+3x\xb) + 
 z(37-26x+14x\xb-11x^2\xb-2x^2-18\xb)
\nonumber \\
\ &+&\
z^2(124x+4x^2\xb-112+x^2+100\xb-48x\xb)
\nonumber \\
\ &+&\ 2z^3(-120+2x^2\xb+5\xb+10x\xb)
+8z^4(2x\xb+32-12x-11\xb)+64z^5]
\nonumber \\
\ &-& \frac{3z}{(1-x)(\xb^2-4z)^2}
[(1-x)(2x-3x\xb+5\xb-2)
\nonumber \\
\ &+& \ z(-6x^2-12x\xb+x^2\xb+28x+16\xb-24)
+ 4z^2(-6+2x+\xb)],
\nonumber \\
b_{24}^{{\rm lc},VV} \ &=&\ b_{13}^{{\rm lc},VV}\Big|_{x\leftrightarrow \xb},
\nonumber \\
b_{14}^{{\rm lc},VV} \ &=& \ 
-\frac{B}{16}\bigg\{
2x^5-3x^4-x^3-40x+x\xb(2x^3+2x^2+5x\xb-43x+36)+8
\nonumber \\ \ &+&\ 8zB[-x^4+7x^3
+8x^2-48x+x\xb(x^3+x^2\xb
-4x^2-3x\xb-x+20)+20]
\nonumber \\
\ &+& \ 64z^2B^2[-3x^3+12x^2-18x
+x\xb(x^2\xb+x^2-8x-x\xb+11)
+5]
\nonumber \\
\ &+&\
128z^3B^2x_g(1-x)^2\bigg\}
\nonumber \\
\ &+&\ \frac{xB}{8(x^2-4z)}
[x^6-8x^4+24x^3-30x^2+22x-6
\nonumber \\
\ &+&\ 
\xb(x^5+4x^4-27x^3+48x^2-44x+12)
+\xb^2(4x^3-17x^2+22x-6)]
\nonumber \\
\ &+&\ \frac{3x^3[x^2+x\xb-2(1-x_g)]^2}{16(1-\xb)(x^2-4z)^2}
+(x\leftrightarrow \xb),\nonumber \\
b_{34}^{{\rm lc},VV} \ &=&\ \frac{B}{8}\bigg\{ 
2x^4+x\xb(2x^2+x-6)-5x^3-2x^2-12x+8
\nonumber \\ \ &+&\  8zB[x\xb(x^2+x\xb-11x+16)
-x^3+12x^2-28x+10
+ 2zx_g^2]
\nonumber \\
\ &-& \ 
\frac{1}{x^2-4z}
[2x^6-2x^5-7x^4+28x^3-36x^2+28x-8
\nonumber \\
\ &+& \ 
2\xb(x^5+2x^4-18x^3+32x^2-30x+8)
+\xb^2(6x^3-25x^2+32x-8)]\bigg\}
\nonumber \\
\ &-&\ 
\frac{3x^2[x^2+x\xb-2(1-x_g)]^2}{8(1-\xb)(x^2-4z)^2}
+(x\leftrightarrow \xb),\nonumber \\
a^{{\rm lc},VV} \ &=&\ \frac{B}{2s}\bigg\{
B[-11x^2+24x+2x\xb(4x-7)-7]+4zB^2[4x^3-2x^2\xb^2-16x^2+23x
\nonumber \\
&+& x\xb(-3x^2+17x-17)-6
-2zx_g(1-x)^2]
- \frac{[x^2+x\xb-2(1-x_g)]^2}
{(2-x)(x^2-4z)} \nonumber \\
&-& \frac{2x^3-8x^2+12x-7+\xb(2-x)^2}
{(2-x)(1-x+z)}
+(x\leftrightarrow \xb)\bigg\},\nonumber \\
k^{{\rm lc},VV} \ &=&\
-\frac{1}{2}x_g^2B\left[\gamma-\ln\left(\frac{4\pi z\mu^2}
{s(1-x)^2}\right)\right]
+ \frac{B}{8}\bigg\{
x^3-16x^2+24x+3x\xb(x-4)-8
\nonumber \\ 
&+& 4zB[5x^3-15x^2 +20x +x\xb(7x-12)-5]
\nonumber \\
 &+& 
16z^2B^2[-4x^3+2x^2\xb^2+16x^2-23x
+ x\xb(3x^2-17x+17)+6
\nonumber \\
&+& 2zx_g(1-x)^2]
-\frac{x[x^2+x\xb-2(1-x_g)]^2}
{x^2-4z}\bigg\} +(x\leftrightarrow \xb).
\eea

\bea
d_{d=6}^{AA} \ &=&\ d_{d=6}^{VV} 
+\frac{2zBs[1-x_g-zBx_g^2]}
{\pi}\bigg\{x^2-10x+x\xb+5 \nonumber \\
\ &+&\ 6zB(2x^2-6x+x\xb+3)+
(x\leftrightarrow \xb)\bigg\},\nonumber \\
c_{123}^{AA} \ &=&\ c_{123}^{VV}-\frac{2zs}{1-x}\bigg\{
x^2-10x+x\xb+5 + 3zBx_g^2
+(x\leftrightarrow \xb)\bigg\},\nonumber \\
c_{124}^{AA} \ &=&\ c_{123}^{AA}\Big|_{x\leftrightarrow \xb},
\nonumber \\
c_{134}^{{\rm lc},AA} \ &=&\ c_{134}^{{\rm lc},VV} + \frac{6zs}{1-x}
\bigg\{-\frac{4z^2}{1-\xb}-
\frac{z(x^2+\xb^2-16(x+\xb)+2x\xb+16)}{3(1-x)}
\nonumber \\
\ &-&\ 
\frac{-(1-x)^2(1-\xb)+2z(1-x)(3x+\xb-9)
+2z^2(2x-\xb-7)}{3(\xb^2-4z)}
\nonumber \\ \ &+&\ \frac{12z^2(2x+\xb-3)}{3(1-x)^2}
- \frac{2z}{(\xb^2-4z)^2}
[(1-x)(x+3\xb-2x\xb-1) \nonumber \\ 
\ &+&\ z(-5x^2+x^2\xb+20x+12\xb-10x\xb-16)
+4z^2(2x+\xb-5)]\bigg\},\nonumber \\
c_{234}^{{\rm lc},AA} \ &=&\ c_{134}^{{\rm lc},AA}\Big|_{x\leftrightarrow \xb},
\nonumber \\
b_{13}^{{\rm lc},AA} \ &=&\ b_{13}^{{\rm lc},VV} -z\bigg\{
\frac{1+z}{1-x}+\frac{1-2z+8z^2}{(1-\xb)(1-4z)}
+\frac{z(1-x)}{(1-\xb)^2}
+
\frac{(1-z)(1-x)-z^2}{(1-x)(1-\xb+z)(1-z)}
\nonumber \\ \ &+&\ 
\frac{1}{(1-x)(\xb^2-4z)(1-z)(1-4z)} 
[-6x^2+14x-2x\xb+\xb-8 \nonumber \\
\ &+& \ 2z(15x^2-31x+5x\xb-4\xb+13)
+2z^2(-12x^2-12x\xb+17\xb+28)
\nonumber \\
\ &+&\ 
4z^3(4x\xb+12x-6\xb-21)+16z^4]
\nonumber \\ \ &+& \ \frac{6}{(1-x)(\xb^2-4z)^2}
[(1-x)^2\xb+z(-6x^2+3x^2\xb+16x-14x\xb+12\xb-10)
\nonumber \\
\ &+&\
2z^2(-x^2+12x-2x\xb+6\xb-16)-8z^3]\bigg\},
\nonumber \\
b_{24}^{{\rm lc},AA} \ &=&\ b_{13}^{{\rm lc},AA}\Big|_{x\leftrightarrow \xb},
\nonumber \\
b_{14}^{{\rm lc},AA} \ &=&\ 
b_{14}^{{\rm lc},VV} +\frac{z}{4}\bigg\{
-\frac{3x^2[x^2+x\xb-2(1-x_g)]^2}{(1-\xb)(x^2-4z)^2}+
B(7x^3-14x^2+13x^2\xb
\nonumber \\
&-& 14x\xb-52x+24) 
- 16zB^2(x^3-13x^2+x^2\xb-8x\xb+33x-14)
\nonumber \\ 
&+& 96z^2B^3x_g(1-x)^2
- \frac{2B}{x^2-4z}[
5x^5-19x^4+34x^3-30x^2+14x-2
\nonumber \\
\ &+&\ 2\xb(2x^2-3x+2)(2x^2-4x+1)
-\xb^2(1-x)(3x^2-6x+2)] + (x\leftrightarrow \xb) \bigg\},
\nonumber \\
b_{34}^{{\rm lc},AA}  \ &=&\ b_{34}^{{\rm lc},VV}+ \frac{z}{2}\bigg\{
\frac{3x[x^2+x\xb-2(1-x_g)]^2}{(1-\xb)(x^2-4z)^2}-
2B(3x^2-20x+3x\xb+10) \nonumber \\
\ &-& \ 8zB^2x_g^2
+ \frac{B}{x^2-4z}[8x^4-29x^3+48x^2-36x+12
\nonumber \\ 
\ &+&\ 
2\xb(6x^3-19x^2+20x-8)
+\xb^2(4x^2-9x+4)] 
+ (x\leftrightarrow \xb) \bigg\}, \nonumber \\
a^{{\rm lc},AA} \ &=&\ a^{{\rm lc},VV}  
+ \frac{1}{s}\bigg\{
-\frac{x^2-x-x\xb+2\xb-1}{(1-x+z)(2-x)(1-\xb)}
\nonumber \\ 
&+& \frac{B}{2}[
x^2-10x+x\xb+3
+4zB(7x^2-19x+4x\xb+8)+24z^2x_gB^2(1-x)^2]
\nonumber \\
\ &+&\ \frac{xB[x^2+x\xb-2(1-x_g)]^2}{2(2-x)(x^2-4z)}
+ (x\leftrightarrow \xb)\bigg\} ,\nonumber \\
k^{{\rm lc},AA}  \ &=&\ k^{{\rm lc},VV} + \frac{zB}{2}
\bigg\{ -2x_g^2\left[\gamma-\ln\left(\frac{4\pi z\mu^2}
{s(1-x)^2}\right)\right]
+\frac{[x^2+x\xb-2(1-x_g)]^2}{x^2-4z}
\nonumber \\
\ &-&\ 10x^2+40x-10x\xb-7
+ 2zB(5x^3-52x^2+15x^2\xb+130x-46x\xb-52)
\nonumber \\
\ &-&\ 24z^2B^2x_g(1-x)^2
+ (x\leftrightarrow \xb) \bigg\}.
\eea
\bea
\tilde{d}_{d=6}^{VV} \ &=&\ 
-\frac{s[1-x_g-zBx_g^2]}
{\pi(1-\xb)(1-x_g-4z)}\bigg\{
x^2+2\xb^2-2x-2\xb+2x\xb+1 \nonumber \\
\ &-&\ \frac{2z}{
(1-\xb)(1-x_g)}[-3x^2-4\xb^2+7x+9\xb-5+x\xb(x+\xb-6) 
\nonumber \\
\ &+&\ 4z(-x^2-3\xb^2+5x+8\xb-3x\xb-6)
-8z^2x_g] \bigg\},\nonumber \\
\tilde{d'}_{d=6}^{VV} \ &=&\ 
\tilde{d}_{d=6}^{VV}\Big|_{x\leftrightarrow \xb},\nonumber \\
\tilde{c}_{123}^{VV} \ &=&\ -\frac{zs}{2}\bigg\{
\frac{2(2-\xb+z)}{1-x} +\frac{2z(1-x)[x(x-4)+4z(-3x+2)+16z^2]}
{(1-\xb)^2(x-4z)x} \nonumber \\
\ &-&\ \frac{(2x-1-4z)[(x-8z)
(2x^2-2x+1)+8z^2(8x-3)-96z^3]}{(1-x)(1-x_g-4z)(x-4z)^2}
\nonumber \\
\ &+&\ \frac{2}{(1-\xb)(x-4z)^2x^2}
[x^3(x^2+x+1)+2zx^2(2x^2-4x-5)
-8z^2x(8x^2-3x-3) \nonumber \\
\ &+&\ 32z^3(5x^2-1)-64z^4(x+1)]
- \frac{x(2x-1)+4z(1+x^2)+8z^2}
{(1-x)(1-x_g)x^2} \bigg\},
\nonumber \\
\tilde{c}_{134}^{VV} \ &=&\ 
\tilde{c}_{123}^{VV}\Big|_{x\leftrightarrow \xb},
\nonumber \\
\tilde{c}_{124}^{VV} \ &=&\ sB(1-x_g-2z)\nonumber 
\\ \ &\times & \ 
\bigg\{x^2+2zB[-3x^2+2x^2\xb-4x\xb+8x-3-zx_g^2]
+  (x\leftrightarrow \xb)\bigg\},\nonumber \\
c_{134}^{{\rm sc},VV} \ &=&\ \tilde{c}_{123}^{VV}-
zs\bigg\{-\frac{z(1-x)(x-8z-4)}
{(1-\xb)^2x}+\frac{x(2x-1)+4z(x^2+1)+8z^2}{(1-x)
(1-x_g)x^2}
 \nonumber \\
\ &+&\ \frac{1}{(1-x)(\xb^2-4z)}[x^2+x^2\xb-6x-5\xb+3x\xb+5
\nonumber \\
\ &+& \ z(x^2-32x-25\xb+10x\xb+48)-8z^2(x+\xb-6)]
\nonumber \\
\ &+&\ \frac{3z}{(1-x)(\xb^2-4z)^2}[
8x^2-5x^2\xb-20x-16\xb+20x\xb+12
\nonumber \\
\ &+&\ 4z(x^2-10x-5\xb+2x\xb+12)+16z^2]
\nonumber \\
\ &-&\ \frac{3\xb(1-x)}{(\xb^2-4z)^2}
-\frac{x(x^2+x+1)
+2z(5x^2-2x-2)-8z^2(x+1)}{(1-\xb)x^2}\bigg\},\nonumber \\
c_{234}^{{\rm sc},VV} \ &=&\ c_{134}^{{\rm sc},VV}\Big|_{x\leftrightarrow \xb},\nonumber \\
\tilde{c}_{234}^{VV} \ &=&\ -\frac{zsB^3}{4}
\bigg\{ \frac{x_g(x-\xb)^2
(x^2+\xb^2+2x+2\xb-3)(1-x)^2}{1-x_g-4z}
\nonumber \\
\ &-&\ \frac{1}{(1-x_g)x_g}[x^8-8x^7+11x^6+90x^5-442x^4
+908x^3-1008x^2+592x-72 \nonumber \\ \ &+&\ 
2x\xb(2x^6-4x^5-80x^4+477x^3-1188x^2+1562x-532)
 \nonumber \\ \ &+&\ 
x^2\xb^2(100x^3-771x^2+2428x-1942)
+ 4x^3\xb^3(-5x^2+67x-148)-17x^4\xb^4]
\nonumber \\
\ &-&\ \frac{4zx_g}{1-x_g}[
x^5-13x^4+56x^3-114x^2+112x-22
\nonumber \\ 
\ &+&\ x\xb(7x^3-58x^2+176x-114)
+ x^2\xb^2(16x-47)] \nonumber \\
\ &+&\ \frac{16z^2x_g^3(1-x)^2}{1-x_g}
+(x\leftrightarrow \xb)\bigg\},\nonumber \\
b_{13}^{{\rm sc},VV} \ &=&\ \frac{1-x-2z}{1-x}
+\frac{z(1+x-4z)}{2(1-\xb)^2}-\frac{z^2(1-x)}{(1-\xb)^3}
\nonumber \\
\ &+&\  \frac{-1+x+2z(1-x)+z^2(x-2)+2z^3}{2(1-\xb+z)(1-z)(1-x)}
\nonumber \\ &-&
\frac{1}{2(1-x)(1-z)(1-4z)(\xb^2-4z)}[(1-x)(1-x+\xb-3x\xb)
+ z(-2x^2
\nonumber \\
&+& 11x\xb(-x+2)-18x-20\xb+25)
+ z^2(x^2+2x\xb(2x-23)+84x+68\xb-60)
\nonumber \\
&+&
2z^3(2x\xb(x+3)+16x+21\xb-124)
+ 4z^4(4x\xb-24x-22\xb+56)+64z^5] 
\nonumber \\
&+& \frac{3z}{(1-x)(\xb^2-4z)^2}[(1-x)(-3x\xb+2x+5\xb-2)
\nonumber \\
&+& z(-6x^2+x\xb(x-12)+28x+16\xb-24)+4z^2(2x+\xb-6)]
\nonumber \\
&+& \frac{B[(1-x)(1+4z(x-3))+2z^2(2x^2-14x+11)+8z^3(2x-1)]}
{2(1-4z)},\nonumber \\
b_{24}^{{\rm sc},VV} \ &=&\ 
b_{13}^{{\rm sc},VV}\Big|_{x\leftrightarrow \xb},
\nonumber \\
b_{14}^{{\rm sc},VV} \ &=&\ -\frac{B(x-\xb)^2(1-x_g)}
{4(1-x_g-4z)}-\frac{3x^3[x^2+x\xb-2(1-x_g)]^2}{16(1-\xb)(x^2-4z)^2}
\nonumber \\
&-& \frac{xB}{8(x^2-4z)}[x^6-10x^4+30x^3-38x^2-6\xb^2
\nonumber \\ &+& x^2\xb^2(4x-17)
+x\xb(x^4+4x^3-29x^2+54x+22\xb-48)+26x+12\xb-6]
\nonumber \\
&+&\frac{B}{16}[2x^5-3x^4-5x^3+5x^2\xb^2+20x^2
+x\xb(2x^3+2x^2-47x+40)-64x+16\nonumber \\
&+& 
8zB(-x^4+5x^3+x^2\xb^2(x-3)+14x^2+x\xb(x^3-4x^2-3x+24)
-56x+22) \nonumber \\ &+&
16z^2B^2(x^4-18x^3+x^2\xb^2(4x-3)+62x^2
+6x\xb(x^2-7x+9)-88x+24)
\nonumber \\ &+&
128x_gz^3B^2(1-x)^2]+ (x\leftrightarrow \xb),\nonumber \\
b_{34}^{{\rm sc},VV} \ &=&\ 
\frac{3x^2[x^2+x\xb-2(1-x_g)]^2}{8(1-\xb)(x^2-4z)^2}+
\frac{B}{8(x^2-4z)}
[2x^6-2x^5-11x^4 \nonumber \\
&+& x^2\xb^2(6x-25)+40x^3-52x^2-8\xb^2
+2x\xb(x^4+2x^3-20x^2+38x+16\xb-34)
\nonumber \\ 
&+& 36x+16\xb-8] 
- \frac{B}{8x_g^2}[2x^6+2x^3\xb^3-13x^5+22x^4+
2x^2\xb^2(3x^2-5x-15)\nonumber \\
&+& 8x^3-40x^2+x\xb(6x^4-25x^3-8x^2+120x-48)
+8 \nonumber \\
&+&8zBx_g^2(x^2\xb^2-x^3+12x^2+x\xb(x^2-11x+15)-26x+9)
+16z^2Bx_g^4] + (x\leftrightarrow \xb),\nonumber \\
\tilde{b}_{24}^{VV} \ &=&\ \frac{B(1-x_g)(x-\xb)^2}
{4(1-x_g-4z)}-\frac{B}{2x_g^2}[x^4+3x^2\xb^2-8x^3+24x^2
\nonumber \\
&+& 
2x\xb(2x^2-12x+11)-28x+6+2zx_g^2]
+ (x\leftrightarrow \xb),\nonumber \\
a^{sc,VV} \ &=&\ 
\frac{B[x^2+x\xb-2(1-x_g)]^2}{2s(2-x)(x^2-4z)}
+ \frac{B[2x^3-8x^2+x\xb(x-4)+12x+4\xb-7]}
{2s(2-x)(1-x+z)} \nonumber \\
&-& \frac{B^3}{2s}[2x^2\xb^2(4x-11)+11x^3-35x^2
+x\xb(-19x^2+71x-45)+38x-7
\nonumber \\
&-& 4z(2x^2\xb^2-4x^3+16x^2+x\xb(3x^2-17x+17)-23x+6)
\nonumber \\
&-& 8z^2x_g(1-x)^2]+ (x\leftrightarrow \xb),\nonumber \\
k^{sc,VV} \ &=&\ \frac{B(1-x_g-2z)x_g^2\ln(\omega)}{2\beta(1-x_g)}
+\frac{Bx[x^2+x\xb-2(1-x_g)]^2}{8(x^2-4z)}
\nonumber \\
&+&\frac{B^2}{8x_g}[B(3x^4\xb^4+x^6+2x^3\xb^3(2x^2-33x+78)
-20x^5+101x^4
\nonumber \\
&+& x^2\xb^2(x^4-28x^3+215x^2-668x+535)-226x^3+272x^2
\nonumber \\
&+&
2x\xb(-x^5+22x^4-128x^3+322x^2-431x+152)-176x+24)
\nonumber \\
&-& 4zx_g(3x^3-7x^2+x\xb(5x-8)+8x-1)
-16z^2Bx_g(2x^2\xb^2-4x^3+16x^2
\nonumber \\
&+&
x\xb(3x^2-17x+17)-23x+6)-32z^3Bx_g^2(1-x)^2]
+ (x\leftrightarrow \xb).
\eea

\bea
\tilde{d}_{d=6}^{AA} \ &=&\ \tilde{d}_{d=6}^{VV} 
-\frac{2sz[1-x_g-zBx_g^2]}
{\pi(1-\xb)(1-x_g)(1-x_g-4z)}\bigg\{
x^3+\xb^3-11x^2-13\xb^2
\nonumber \\ 
&+&
x\xb(3x+3\xb-22)+20x+24\xb-12
-
\frac{2z}{1-\xb}[
-\xb^3+7x^2+21\xb^2 \nonumber \\ 
&+& 
x\xb(-x-2\xb+24)-28x-46\xb+26
+12zx_g]\bigg\},\nonumber \\
\tilde{d'}_{d=6}^{AA} \ &=&\ 
\tilde{d}_{d=6}^{AA}\Big|_{x\leftrightarrow \xb},\nonumber \\
\tilde{c}_{123}^{AA} \ &=&\ \tilde{c}_{123}^{VV} +zs\bigg\{
-\frac{2[x+\xb-3+z(x-3)-z^2]}{1-x}
-\frac{x^3+zx(4x+1)+12z^2}{(1-x)(1-x_g)x^2}\nonumber \\
&+& \frac{(1-2x+4z)[x^3-zx(8x+1)+8z^2(6x-1)-16z^3(x+5)+64z^4]}
{(1-x)(1-x_g-4z)(x-4z)^2} \nonumber \\
&-& \frac{2z}{(1-\xb)(x-4z)^2x^2}
[x^5-15x^4+x^3+2z(-2x^4+52x^3+3x^2)
\nonumber \\
&+&
8z^2(-x^3-28x^2-5x)
+ 32z^3(x^2+3x+3)]
- \frac{2z^2(1-x)[x^2+12x-4z(x+6)]}{(1-\xb)^2(x-4z)x}
\bigg\},\nonumber \\
\tilde{c}_{134}^{AA} \ &=&\ 
\tilde{c}_{123}^{AA}\Big|_{x\leftrightarrow \xb},\nonumber \\
\tilde{c}_{124}^{AA} \ &=&\ \tilde{c}_{124}^{VV}
+2zsB\bigg\{(1-x_g)(x^2+x\xb-10x+5) 
-2zB[x^2\xb^2-4x^3+26x^2
\nonumber \\ 
&+& x\xb(x^2-22x+31)-44x+11]
-6z^2Bx_g^2+ (x\leftrightarrow \xb)\bigg\},\nonumber \\
c_{134}^{{\rm sc},AA} \ &=&\ c_{134}^{{\rm sc},VV} 
+ \tilde{c}_{123}^{AA} - \tilde{c}_{123}^{VV} +
2zs\bigg\{\frac{z^2(1-x)(x+12)}{(1-\xb)^2x}
+\frac{x^3+zx(4x+1)+12z^2}{(1-x)(1-x_g)x^2}
\nonumber \\
&-& \frac{x^2+x\xb(1-x)-2x+1+2z(3x^2-9x+2\xb+5)
+2z^2(-2x+\xb+3)}{(1-x)(\xb^2-4z)} \nonumber \\
&+& \frac{6z}{(1-x)(\xb^2-4z)^2}[-x^2+x\xb(2x-5)+2x+3\xb-1
+z(-5x^2
\nonumber \\
&+&
x\xb(x-10)+20x+12\xb-16)
+ 4z^2(2x+\xb-5)] \nonumber \\
&+& \frac{z[x^3-15x^2+x+2z(x^2+6x+6)]}{(1-\xb)x^2}
+\frac{x+\xb-3+z(2x+\xb-6)+z^2}{1-x}\bigg\},\nonumber \\
c_{234}^{{\rm sc},AA} \ &=&\ 
c_{134}^{{\rm sc},AA}\Big|_{x\leftrightarrow \xb},\nonumber \\
\tilde{c}_{234}^{AA} \ &=&\ \tilde{c}_{234}^{VV} 
+\frac{zsB^3x_g}{4}\bigg\{-\frac{1}{1-x_g}[
2x^6+x^3\xb^3(-2x-13)-6x^5-12x^4
\nonumber \\
&+& x^2\xb^2(x^3-22x^2+176x-250)
+ 98 x^3-194x^2+x\xb(x^5-7x^4+54x^3
\nonumber \\
&-& 258x^2+542x-246)
+ 160x -24 
- 4z(3x^3\xb^3-4x^5+42x^4
\nonumber \\
&+&
x^2\xb^2(4x^2-66x+144)-148x^3
+262x^2
+ x\xb(x^4-34x^3+190x^2-476x+274)
\nonumber \\
&-&
232x+40)-48z^2x_g^2(1-x)^2]
+\frac{(x-\xb)^2}{1-x_g-4z}[2x^4+x^2\xb^2(3x-3)
\nonumber \\
&-& 6x^3
+12x^2+x\xb(x^3-3x^2-6x+11)-14x+3]
+ (x\leftrightarrow \xb)\bigg\},\nonumber \\
b_{13}^{{\rm sc},AA} \ &=&\ b_{13}^{{\rm sc},VV} 
+z\bigg\{\frac{3-z}{1-x}+\frac{(1-x)(1-z)-z^2}{(1-x)(1-z)(1-\xb+z)}
+\frac{z(1+x-4z)}{(1-\xb)^2}-\frac{2z^2(1-x)}{(1-\xb)^3}
\nonumber \\
&-& \frac{B[(1-x)(-5+22z)+2z^2(12x-11)-8z^3]}{1-4z}
\nonumber \\ 
&-&
\frac{1}{(1-x)(1-z)(1-4z)(\xb^2-4z)}
[6x^2-2x\xb-10x+5\xb+4
\nonumber \\
&-& 2z(15x^2-5x\xb-23x+9\xb+11)+2z^2(12x^2+4x\xb-2x-15\xb+4)
\nonumber \\ 
&-& 4z^3(4x\xb+8x-10\xb-5)-16z^4]
-\frac{6}{(1-x)(\xb^2-4z)^2}[x\xb(-x+2)-\xb \nonumber \\
&-& z(-6x^2+x\xb(3x-14)+16x+12\xb-10)
\nonumber \\
&+& 2z^2(x^2+2x\xb-12x-6\xb+16)+8z^3] \bigg\}, \nonumber \\
b_{24}^{{\rm sc},AA} \ &=&\ 
b_{13}^{{\rm sc},AA}\Big|_{x\leftrightarrow \xb},
\nonumber \\
b_{14}^{{\rm sc},AA} \ &=&\ b_{14}^{{\rm sc},VV}
+\frac{z}{4}\bigg\{\frac{4B(x-\xb)^2}{1-x_g-4z}
+\frac{3x^2[x^2+x\xb-2(1-x_g)]^2}{(1-\xb)(x^2-4z)^2}
\nonumber \\
&+& \frac{2B}{x^2-4z}[5x^5-21x^4+3x^2\xb^2(x-3)+40x^3-38x^2-2\xb^2
\nonumber \\
&+& 2x\xb(4x^3-15x^2+21x+4\xb-13)+18x+4\xb-2]
-B^2[B(x^3\xb^3(13x-44)
\nonumber \\ &+& 7x^5-32x^4 +
x^2\xb^2(7x^3-58x^2+128x+20)-9x^3+150x^2
\nonumber \\
&+& 
x\xb(-14x^4+77x^3-44x^2-267x+214)
- 180x+32)-8z(x^3-21x^2
\nonumber \\
&-&x\xb(x+10)+56x-25)+ 8z^2Bx_g(x+\xb+10)(1-x)^2]
+ (x\leftrightarrow \xb)\bigg\},\nonumber \\
b_{34}^{{\rm sc},AA} \ &=&\ b_{34}^{{\rm sc},VV}+\frac{z}{2}
\bigg\{\frac{-3x[x^2+x\xb-2(1-x_g)]^2}{(1-\xb)(x^2-4z)^2}
-\frac{B}{x^2-4z}[8x^4-33x^3
\nonumber \\
&+& 4x^2\xb^2+60x^2+4\xb^2
+ x\xb(12x^2-42x-9\xb+52)-52x-24\xb+20]
\nonumber \\
&+&
2B^2[3x^2\xb^2-3x^3+23x^2
\nonumber \\
&+&
x\xb(3x^2-29x+33)-40x+10 + 4zx_g^2]
+ (x\leftrightarrow \xb)\bigg\},\nonumber \\
\tilde{b}_{24}^{{\rm sc},AA} \ &=&\ \tilde{b}_{24}^{{\rm sc},VV}
-\frac{2zB(x-\xb)^2}{1-x_g-4z},\nonumber \\
a^{{\rm sc},AA} \ &=&\ a^{{\rm sc},VV}+\frac{B}{2s}\bigg\{
-\frac{x[x^2+x\xb-2(1-x_g)]^2}{(x^2-4z)(2-x)}
+\frac{2(1-x)(x^2-x\xb-x+2\xb-1)}{(1-x+z)(2-x)}
\nonumber \\
&-& B[B(x^3\xb^3+x^4+x^2\xb^2(x^2-16x+28)-12x^3+27x^2
\nonumber \\
&+& x\xb(-2x^3+26x^2-68x+33)-22x+3) 
\nonumber \\ 
&+&
4z(7x^2+4x\xb-19x+8)
+ 24z^2Bx_g(1-x)^2]
+(x\leftrightarrow \xb)\bigg\},\nonumber \\
k^{sc,AA} \ &=&\ k^{sc,VV}+ \frac{z}{2}\bigg\{
\frac{2B(1-x_g-2z)x_g^2\ln(\omega)}{\beta(1-x_g)}
-\frac{B[x^2+x\xb-2(1-x_g)]^2}{(x^2-4z)}
\nonumber \\
&-& B^2[B(-10x^3\xb^3-10x^4+x^2\xb^2(-10x^2+104x-149)
+64x^3-120x^2\nonumber \\
&+& 2x\xb(10x^3-74x^2+162x-71)+88x-11)
+2z(3x^3-40x^2
\nonumber \\
&+&
x\xb(9x-34)+106x-44)-24z^2Bx_g(1-x)^2]
+(x\leftrightarrow \xb)\bigg\}.
\eea

\newpage
  \begin{figure}
    \begin{center}
      \epsfig{file=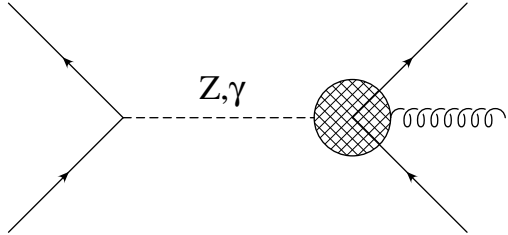}
      \caption{Amplitude for the process $e^+(p_+)e^-(p_-)
        \rightarrow Z^\ast,\gamma^\ast\rightarrow Q(k_1) 
        \bar Q(k_2)g(k_3)$}
      \label{fig:amplitude}
    \end{center}
  \end{figure}
  \begin{figure}
    \begin{center}
      \epsfig{file=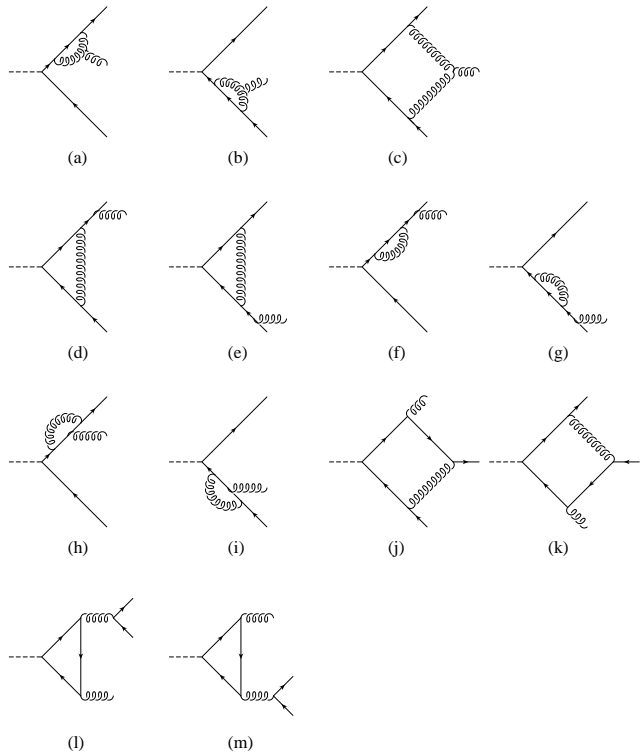,height=10cm}
      \caption{Loop diagrams.}
      \label{fig:loop}
    \end{center}
  \end{figure}
  \begin{figure}
    \begin{center}
      \epsfig{file=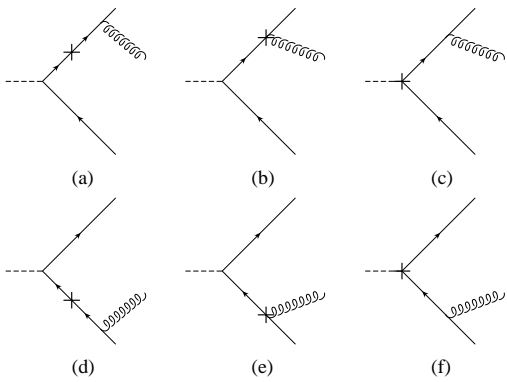,height=5cm}
      \caption{Counterterm diagrams.}
      \label{fig:counter}
    \end{center}
  \end{figure}

\begin{figure}
\begin{center}
\epsfig{file=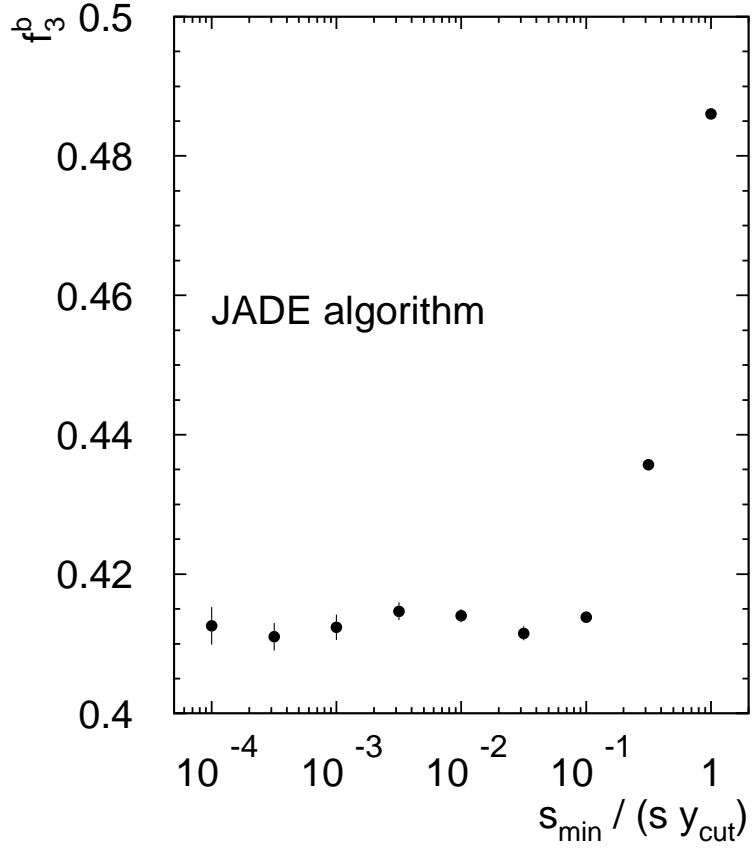,height=12cm}
\caption{
The three jet fraction 
$f_3^b$ at NLO as defined in the text at 
$\protect\sqrt{s}=\mu=m_Z$ as a function of 
$y_{min}=s_{min}/(sy_{cut})$
for the JADE algorithm at a value of the jet resolution
parameter $y_{cut}=0.03$ with 
$m^{\scriptsize{\mbox{$\msbar$}}}_b(\mu=m_Z)=3$ GeV and 
$\alpha_s(\mu=m_Z)=0.118$.
}
\label{fig:sminja}
\end{center}
 \end{figure}

\begin{figure}
 \begin{center}
\epsfig{file=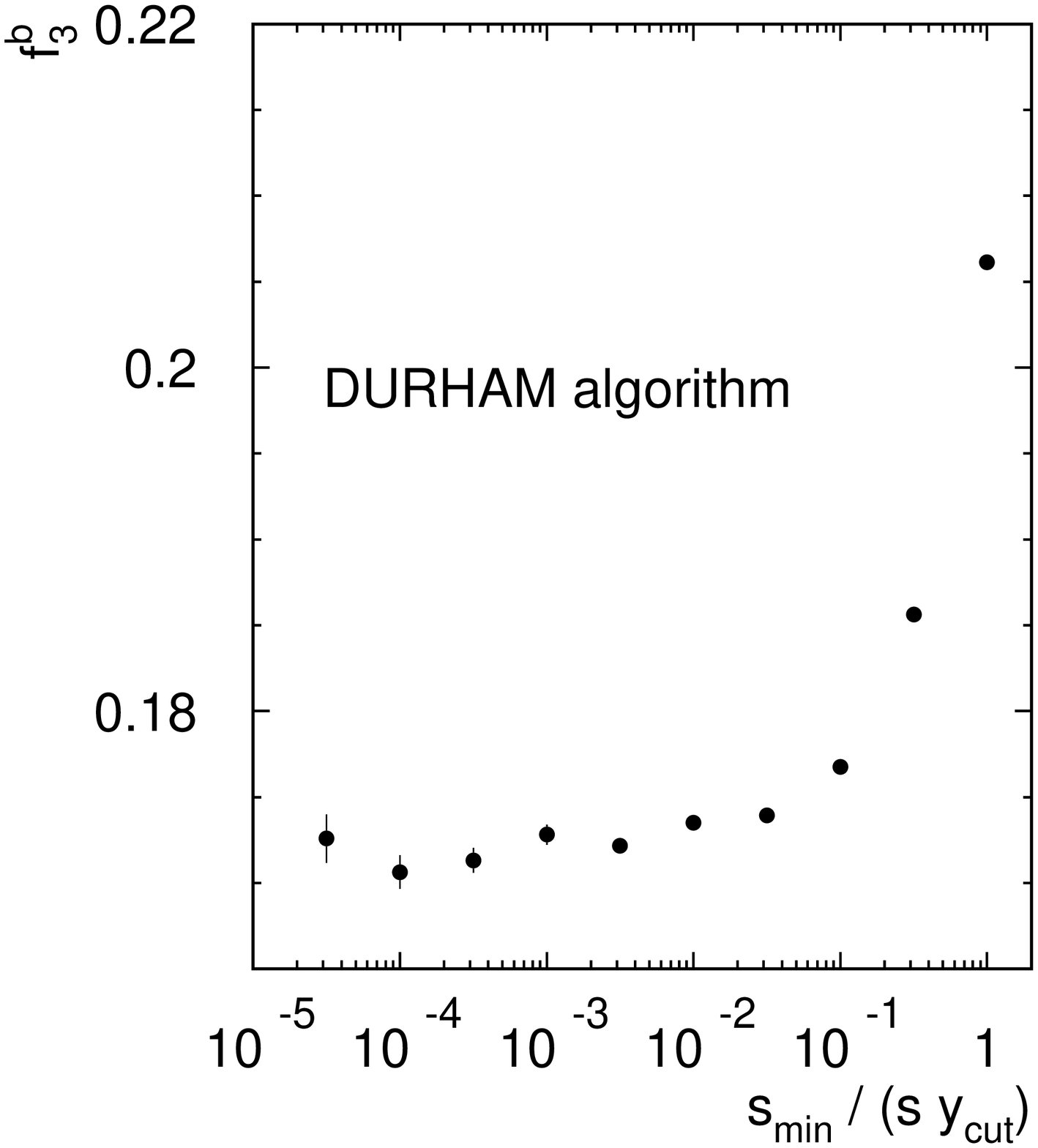,height=12cm}
\caption{
Same as Fig. \ref{fig:sminja}, but for the Durham algorithm.
}
\label{fig:smindu}
\end{center}
 \end{figure}

\begin{figure}
 \begin{center}
\epsfig{file=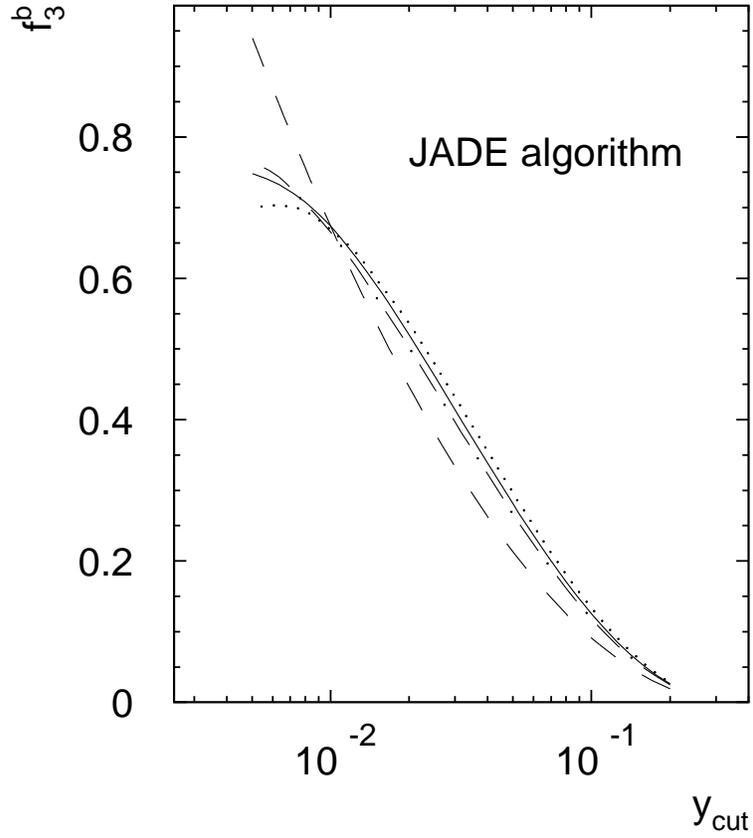,height=12cm}
\caption{
The three jet fraction 
$f_3^b$ as defined in the text at 
$\protect\sqrt{s}=m_Z$
as a function of $y_{cut}$ for the JADE algorithm. 
The dashed line is the LO result.
The NLO results are for $\mu=m_Z$ (solid line),
$\mu=m_Z/2$ (dotted line), and $\mu=2m_Z$ (dash-dotted
line). 
}
\label{fig:ycutja}
\end{center}
 \end{figure}
\begin{figure}
 \begin{center}
\epsfig{file=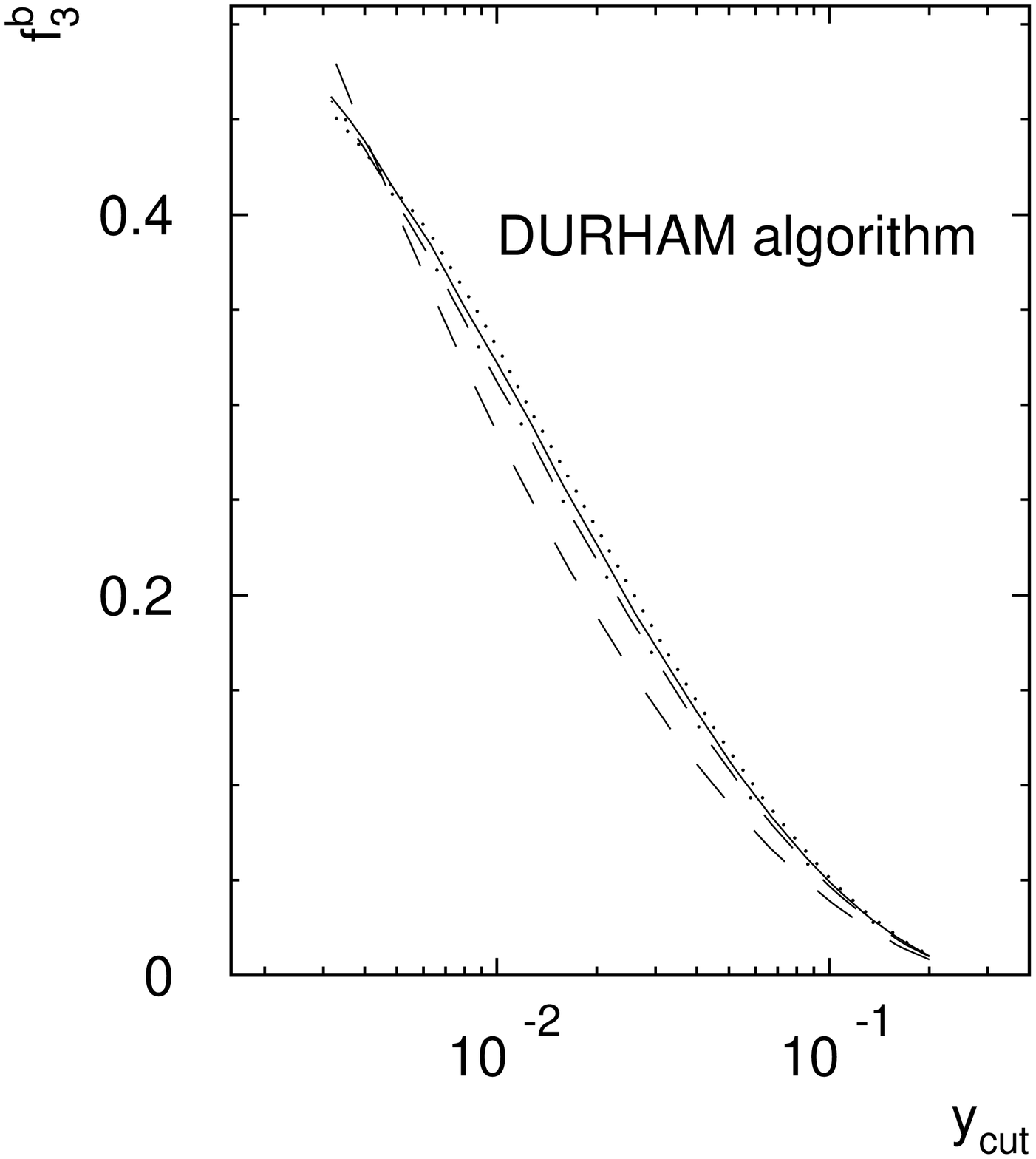,height=12cm}
\caption{
Same as Fig. \ref{fig:ycutja}, but for the Durham algorithm.
}
\label{fig:ycutdu}
\end{center}
 \end{figure}
\begin{figure}
 \begin{center}
\epsfig{file=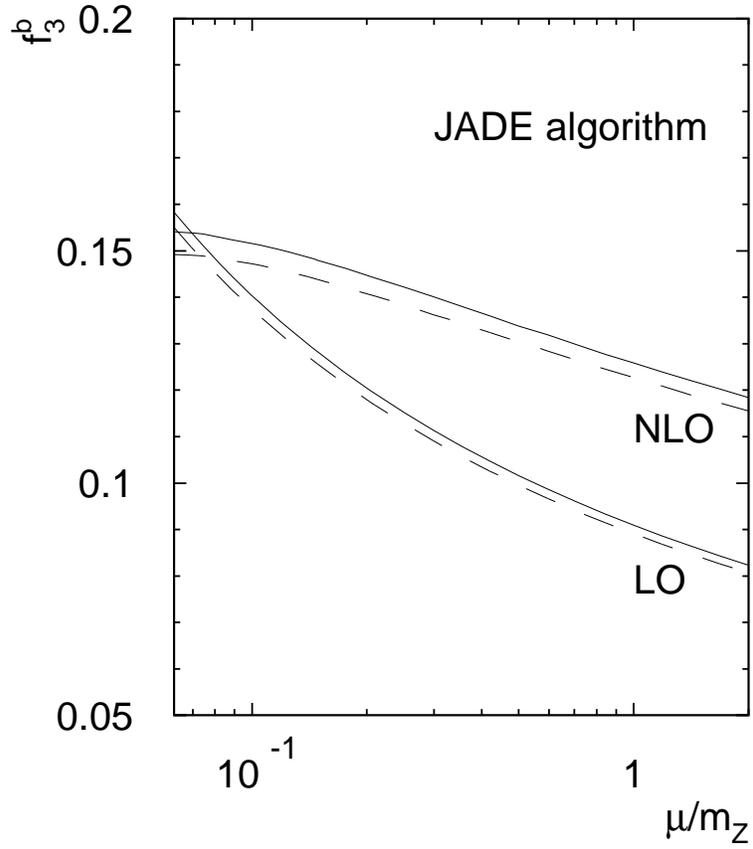,height=12cm}
\caption{Dependence of the three jet fraction $f_3^b$ for
the JADE algorithm
on the renormalization scale at $\protect\sqrt{s}=m_Z$ 
and $y_{cut}=0.2\times 10^{-3/10}\approx 0.1$
for on-shell $b$ quark masses $m_b^{\rm pole}=3$ GeV 
(full curves) and
$m_b^{\rm pole}=5$ GeV (dashed curves).
}
\label{fig:renormja}
\end{center}
\end{figure}
\begin{figure}
 \begin{center}
\epsfig{file=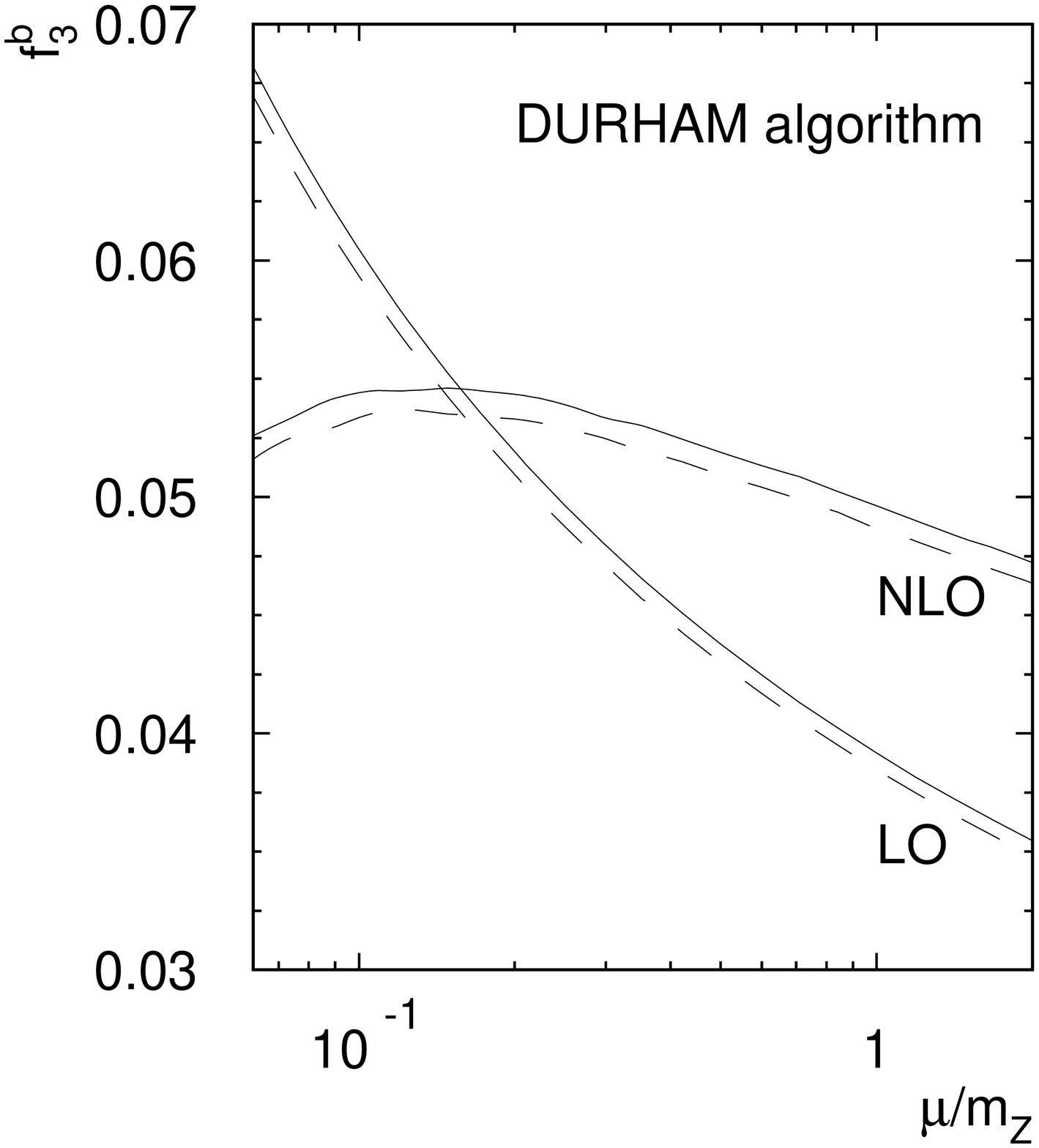,height=12cm}
\caption{
Same as Fig. \ref{fig:renormja}, but for the Durham algorithm.
}
\label{fig:renormdu}
\end{center}
 \end{figure}
\begin{figure}
 \begin{center}
\epsfig{file=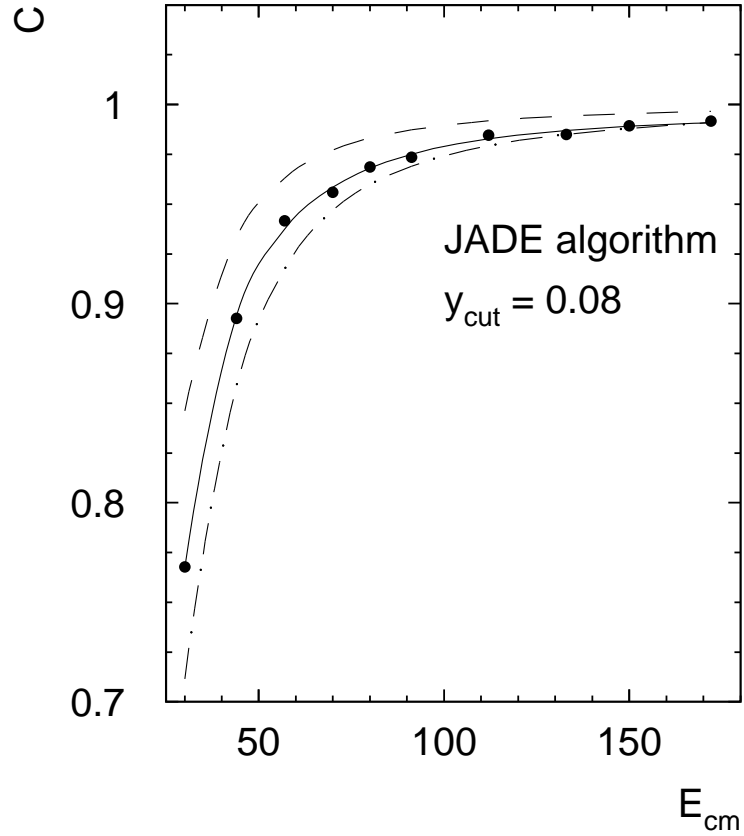,height=12cm}
\caption{The double ratio ${\cal C}$ 
as a function of the c.m. energy 
at $y_{cut}=0.08$ for the JADE algorithm.
The full curve and the points show the NLO result, the dashed curve shows the LO 
result. In both cases a running
mass $m^{\scriptsize{\mbox{$\msbar$}}}_b(\mu=\protect\sqrt{s})$
evolved from $m^{\scriptsize{\mbox{$\msbar$}}}_b(\mu=m_Z)=3$ GeV is used.
The dash-dotted curve shows the LO result for a fixed  mass 
$m_b=4.7$ GeV.}
\label{fig:edep}
\end{center}
 \end{figure}
\begin{figure}
 \begin{center}
\epsfig{file=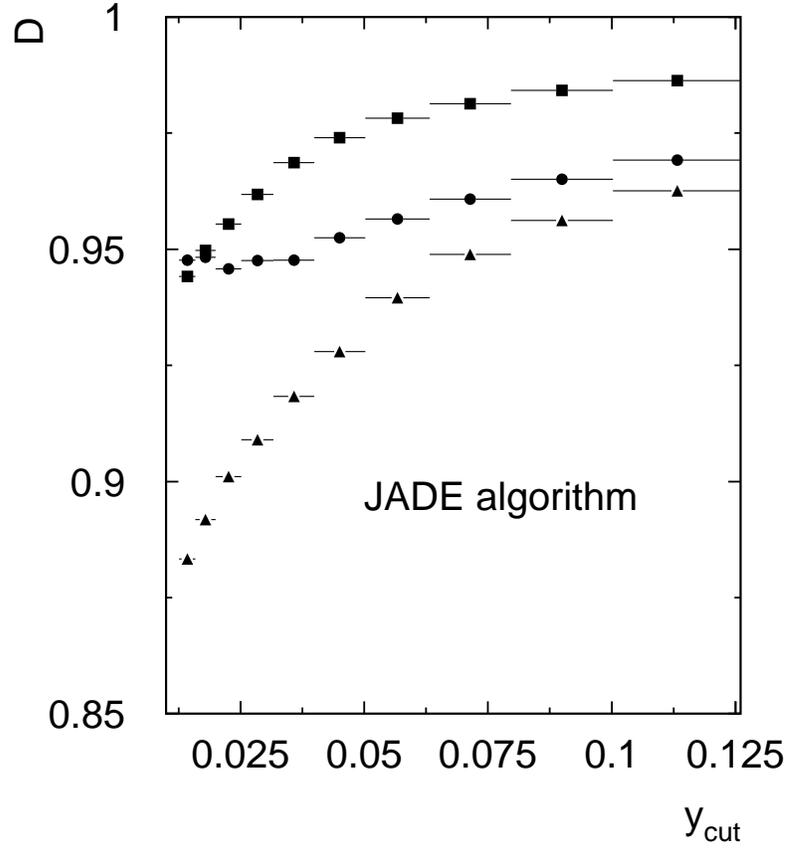,height=12cm}
\caption{The double ratio ${\cal D}$ as a function
of $y_{cut}$
for $\protect\sqrt{s}=\mu=m_Z$.
Full circles: NLO results for 
$m^{\scriptsize{\mbox{$\msbar$}}}_b(\mu=m_Z)=3$ GeV. 
For comparison, the
squares (triangles) are the LO results for
$m_b=3$ GeV ($m_b=5$ GeV). The horizontal bars
show the size of the bins in $y_{cut}$.
}
\label{fig:d2ja}
\end{center}
\end{figure}
\begin{figure}
 \begin{center}
\epsfig{file=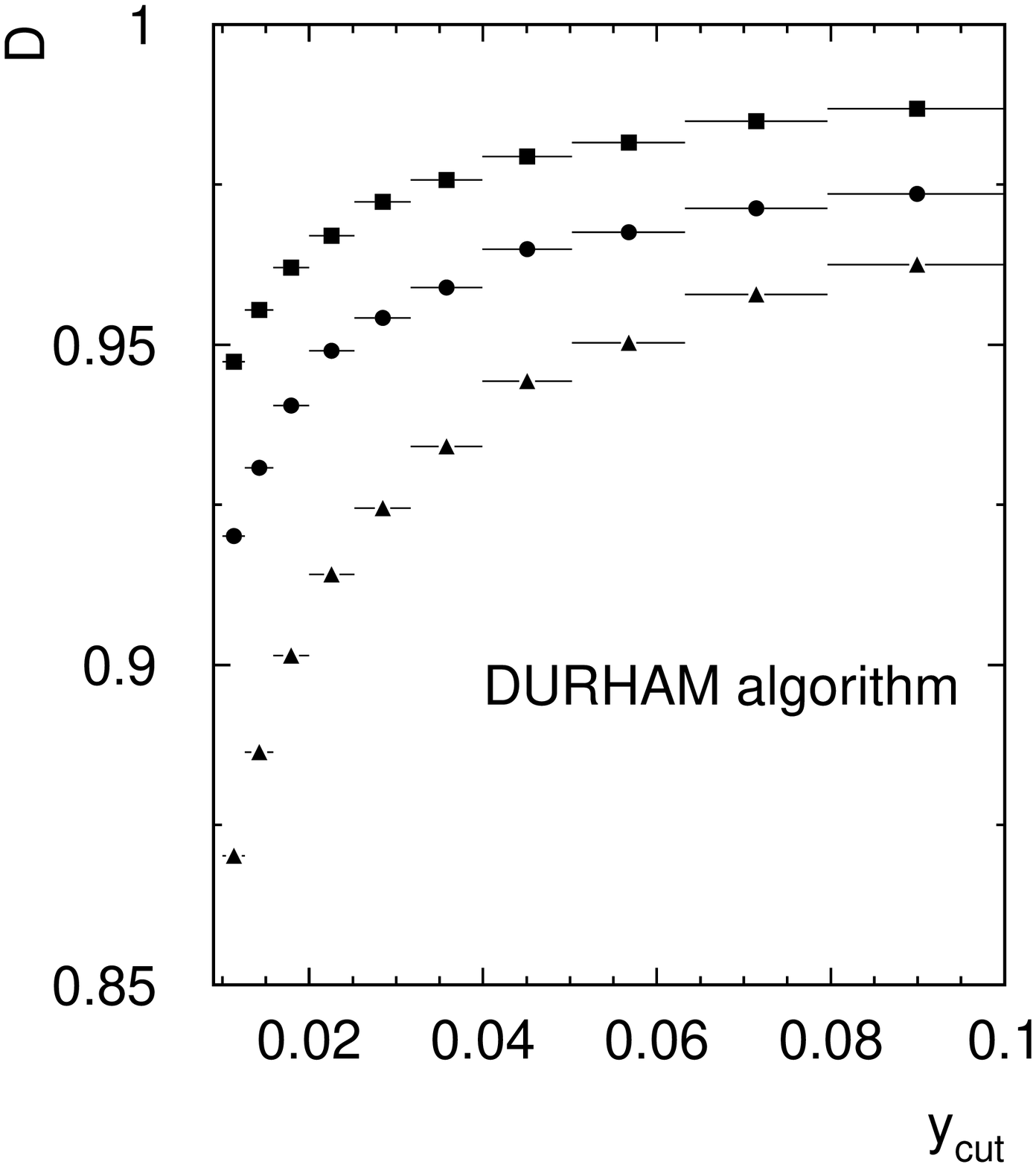,height=12cm}
\caption{Same as Fig. \ref{fig:d2ja}, but for the Durham algorithm.
}
\label{fig:d2du}
\end{center}
\end{figure}
\begin{figure}
 \begin{center}
\epsfig{file=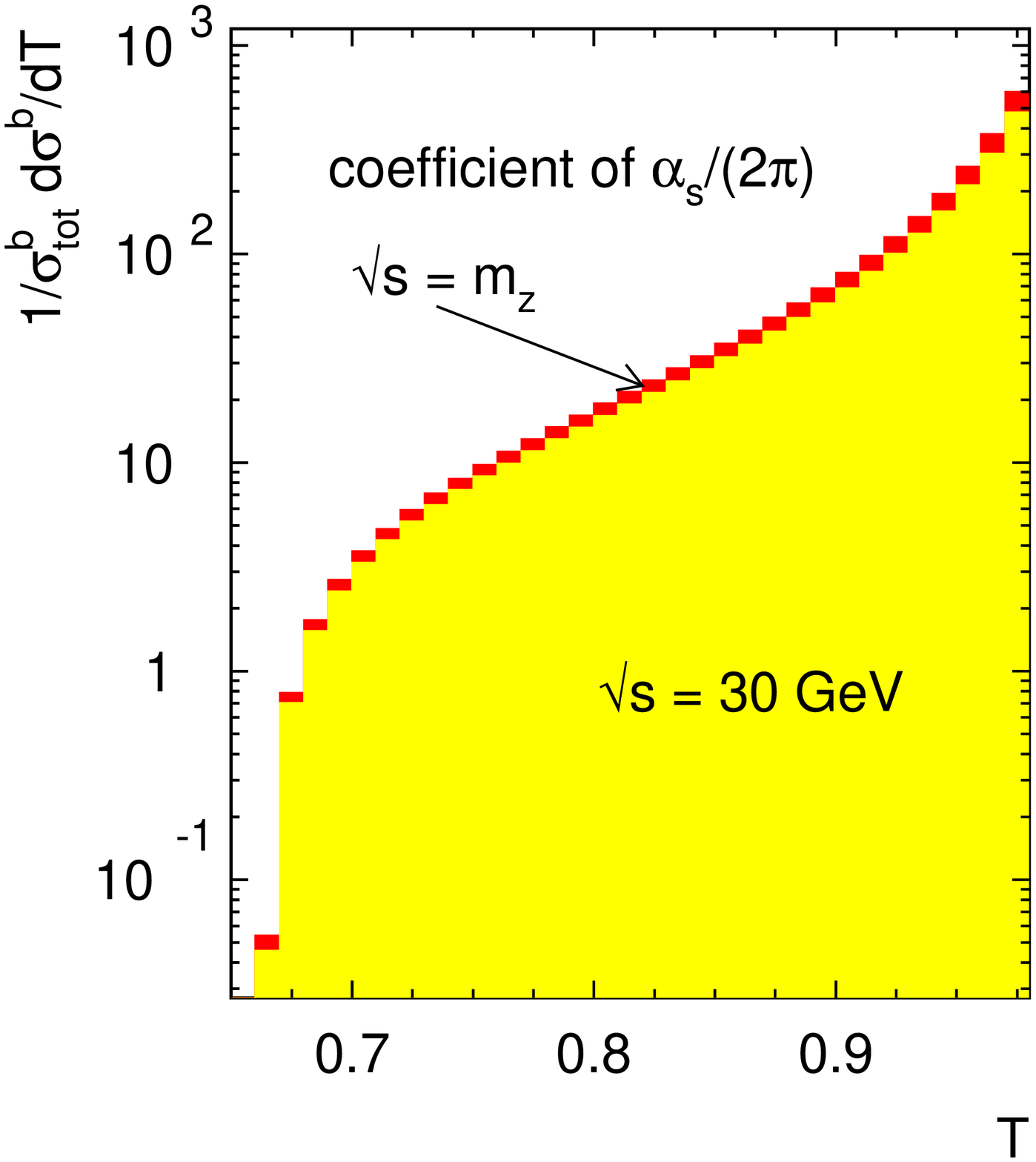,height=12cm}
\caption{Coefficient $c_1$ of the thrust distribution (\ref{thrdist}) 
for $b$ quarks
at $\protect\sqrt{s}=m_Z$ (upper curve) and $\protect\sqrt{s}=30$ GeV (lower curve).}
\label{fig:thrLO}
\end{center}
\end{figure}
\begin{figure}
 \begin{center}
\epsfig{file=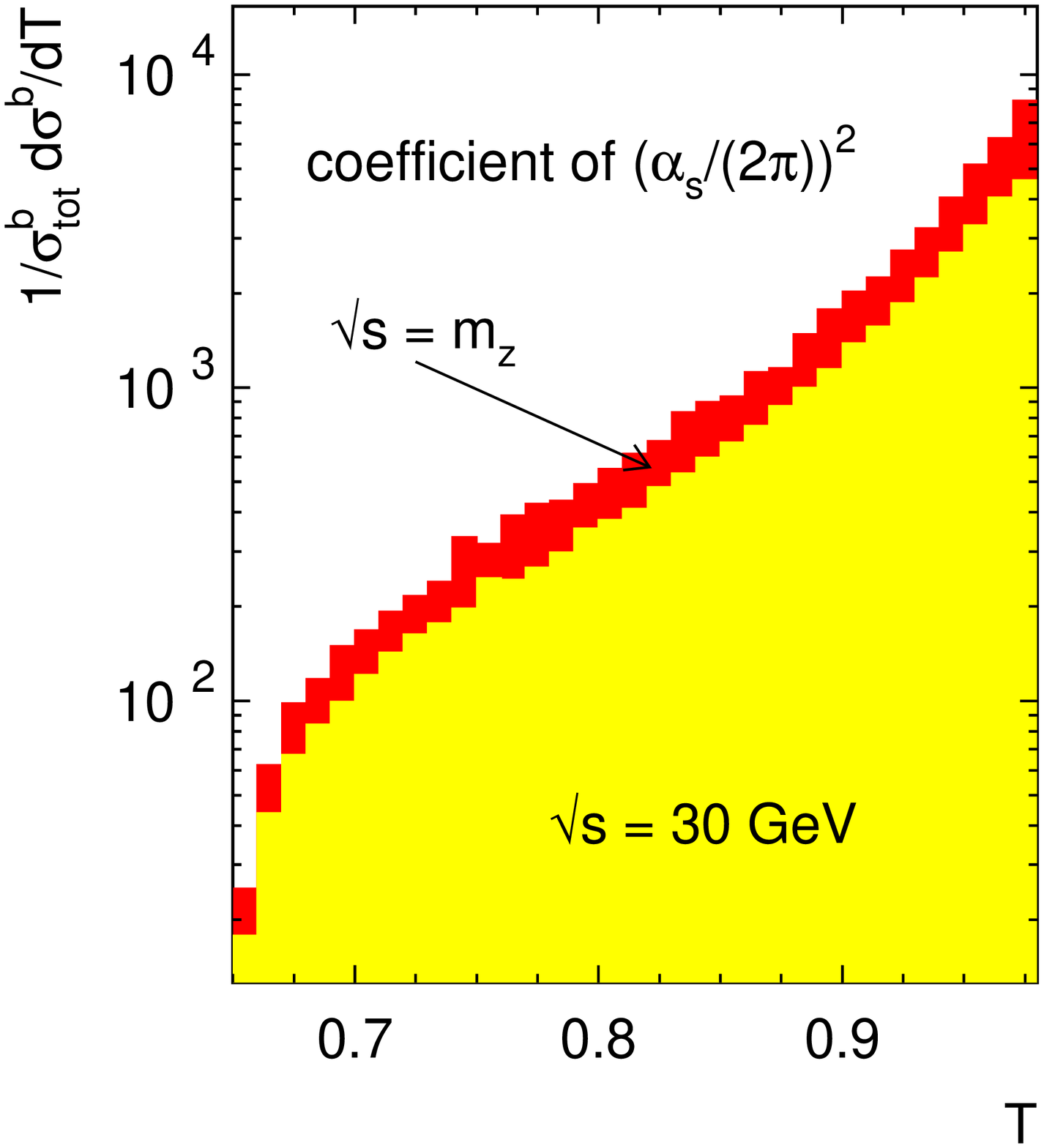,height=12cm}
\caption{Coefficient $c_2$ of the thrust distribution (\ref{thrdist}) 
for $b$ quarks
at $\protect\sqrt{s}=m_Z$ (upper curve) and $\protect\sqrt{s}=30$ GeV (lower curve).
}
\label{fig:thrNLO}
\end{center}
\end{figure}
\begin{figure}
 \begin{center}
\epsfig{file=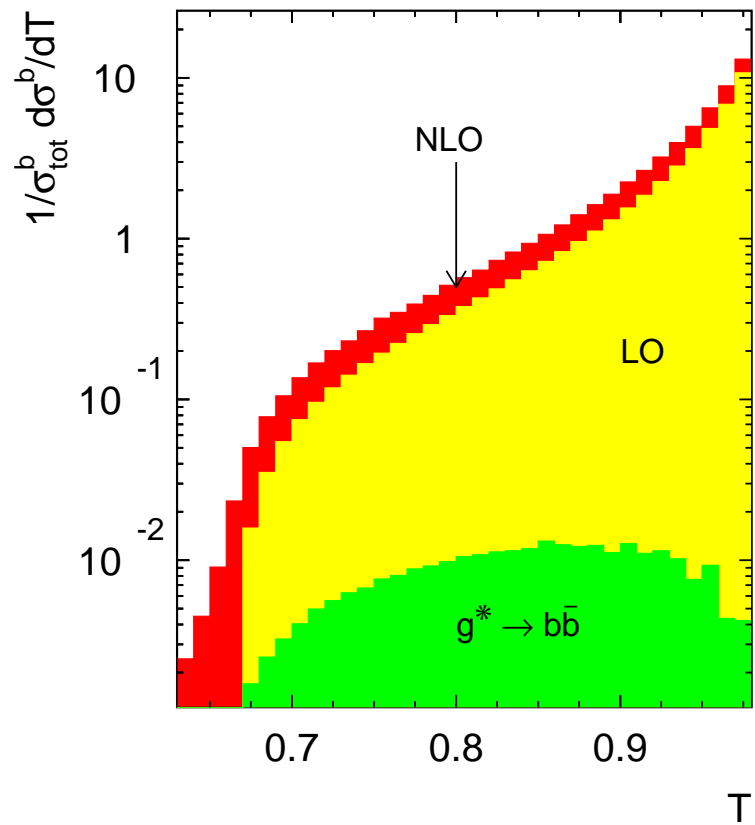,height=12cm}
\caption{The thrust distribution (\ref{thrdist}) at $\protect\sqrt{s}=30$ GeV. Shown separately
is the contribution from the process $e^+e^-\to q\bar{q}g^{\ast}\to q\bar{q}b\bar{b}$.
}
\label{fig:thr30}
\end{center}
\end{figure}
\begin{figure}
 \begin{center}
\epsfig{file=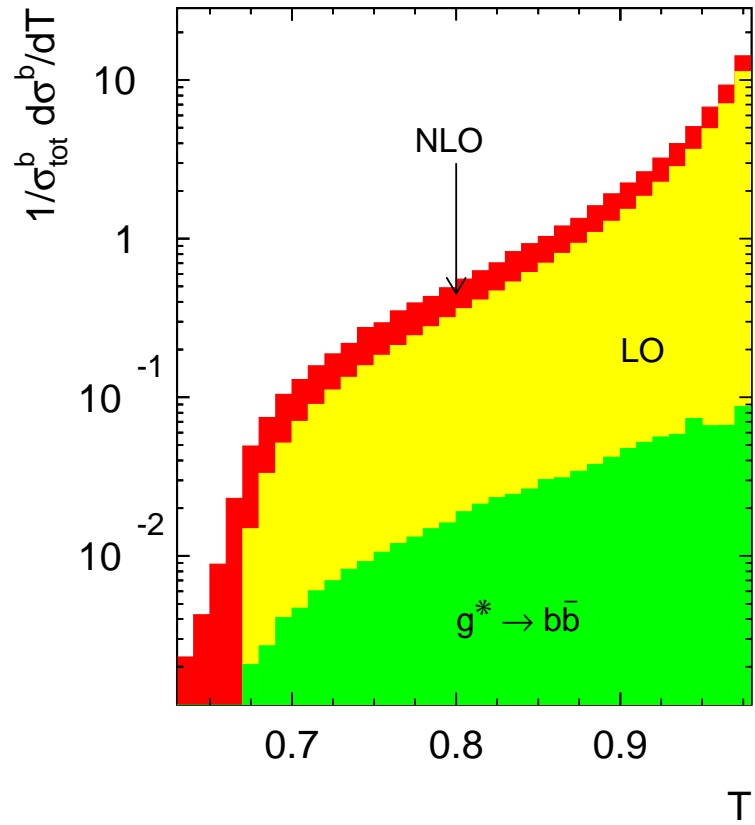,height=12cm}
\caption{Same as Fig. \ref{fig:thr30}, but at $\protect\sqrt{s}=m_Z$.
}
\label{fig:thr91}
\end{center}
\end{figure}
\end{document}